# Probing functionalities and acidity of calcined phenylene-bridged periodic mesoporous organosilicates using (DNP)-NMR, DRIFTS and XPS

*Cyril Pirez[a]\*, Hiroki Nagashima[a,b], Franck Dumeignil[a], Olivier Lafon[a,c]\**

[a] Univ. Lille, CNRS, Centrale Lille, ENSCL, Univ. Artois, UMR 8181 – UCCS – Unité de Catalyse et Chimie du Solide, F-59000 Lille, France

[b] Interdisciplinary Research Center for Catalytic Chemistry, National Institute of Advanced Industrial Science and Technology (AIST), 1-1-1 Higashi, Tsukuba, Ibaraki 305-8565, Japan

[c] Institut Universitaire de France (IUF), 1 rue Descartes, 75231 Paris, France

ABSTRACT Owing to their high surface area, their high stability and their hydrophobicity, periodic mesoporous organosilica (PMO) materials represent promising catalytic support for environmental-friendly chemical processes in water. We investigate here how the calcination of PMO material with benzene linkers (PMOB) allows its functionalization. Conventional and Dynamic Nuclear Polarization (DNP)-enhanced NMR spectroscopy, diffuse reflectance infrared Fourier transform spectroscopy and X-ray photoelectron spectroscopy prove that calcination at 450°C results in the oxidation of phenylene bridges into (poly)phenols but also in carboxylic acids. Ketone, aldehyde as well as allyl and aliphatic alcohol functionalities are also observed but their amount is much lower than that of carboxylic acids. The calcination also cleaves the Si-C bonds. Nevertheless, $N_2$ adsorption-desorption measurements, powder X-ray diffraction and transmission electron microscopy indicates that the PMOB materials calcined up to 600°C still exhibit ordered mesopores. We show that the phenol and carboxylic acid functionalities of PMOB calcined at 450°C protonate the $NH_2$ group of 1-(3-aminopropyl)imidazole (API) in water at room temperature but no formation of covalent bond between API and the calcined PMOB functionalities have been detected.

1. **Introduction**

Green chemistry is today a major scientific and societal challenge. The development of environmental-friendly chemical processes requires that chemists inspire from chemical reactions occurring in living organisms. Nature has notably evolved efficient and selective catalysts to control chemical transformation. However, the cost of enzymes, their inherent instability and their restricted operating window limit their application in industry. In such context, it is highly desirable to design simpler synthetic catalysts that are inspired from enzyme and can perform similar chemistry. A key feature of enzyme is the hydrophobicity of the active site nanoenvironment, which increases the affinity of the hydrophobic substrates for the active site in water.[1]

Building on this bio-inspiration, an opportunity in catalyst design is to develop mesoporous organic materials, where organic active sites are present on the surface. In particular, owing to their high surface area, their high stability and their hydrophobicity, periodic mesoporous organosilica (PMO) materials represent promising supports for the synthesis of these catalysts. PMO materials have been discovered, in 1999, by three different research groups. They reported the possibility to condensate organosilicon precursors in order to create highly ordered mesoporous hydrophobic materials.[2–4] Since this discovery, PMO materials have already been reported with different organic bridging moieties, like methylene[5–7], ethylene[4,7,8], ethylidene[3,9–11], phenylene[7,12–17], biphenylene[18,19], thiophene[12,20,21], anthracene[22–24], and even chiral organic groups[25]. The diversity of organic linkers has led to a vast chemical versatility of these materials and the possibility of fine-tuning of their chemical environment. Benzene bridging materials are the most reported materials in the literature for catalysis application because they benefit from homogeneous distribution of organic moieties, thermal stability, hydrophobicity, and crystallinity.[13,14,16] PMO materials with benzene linkers (PMOB) have already been used to immobilize enzymes[26] and organometallics[27] using hydrophobic interactions. Furthermore, the hydrophobicity of PMOB materials can facilitate the elimination of water from the vicinity of the active site for reactions, for which water is an undesirable product.[28] This material can also preferentially adsorb hydrophobic molecules rather than hydrophilic ones.[29] Recently, researchers studied the mechanic stability of PMOB material under calcination using different gases. They showed that it was possible to generate quinoid species on the surface of PMOB material.[30]

We study here the controlled calcination of PMOB materials in order to generate carbonyl groups, and carboxylic ones in particular, on the surface of their pores while keeping high specific area, their 2D hexagonal structure and the benzene ring arrangement conferring a suitable hydrophobic core. Furthermore, we study the reactivity between these surface carbonyl groups and amines in water at room temperature without the use of reagent to create strong interaction between the acid moieties present on the silica surface and the basic moieties from the amine compound. The PMOB are characterized using thermogravimetry analysis (TGA), $N_2$ adsorption-desorption measurements, powder X-ray diffraction (XRD), transmission electron microscopy (TEM), diffuse reflectance infrared Fourier transform spectroscopy (DRIFTS), conventional $^{29}$Si and $^{13}$C NMR spectroscopy, dynamic nuclear polarization (DNP)-enhanced $^{13}$C NMR spectroscopy and X-ray photoelectron spectroscopy (XPS). DNP can enhance the sensitivity of NMR spectroscopy by one to three orders of magnitude through the microwave-driven transfer of polarization from unpaired electrons to the surrounding nuclei.[31–33] Such technique has notably been employed to probe the atomic-level structure of material surfaces.[34–36] Nevertheless, to the best of our knowledge, DNP-NMR spectroscopy of PMO has been limited to the characterization of phenylpyridine-based PMO and its organometallic derivatives.[37] To the best of our knowledge, this technique has not yet been applied to study the defects formed during the calcination of PMOs.

## 2. Experimental Section

**Products.** We employed the following chemicals (the supplier being indicated within parentheses): HCl 36wt% (Fisher), ethanol (Fisher, 99%), 1.4-bis(triethoxysilyl)benzene (BTSB, Sigma-Aldrich, 96%), 1-(3-Aminopropyl)imidazole (API) (Sigma-Aldrich, ≥97%), tetraethyl orthosilicate (Sigma-Aldrich, 99%), Pluronic® P-123 (Aldrich).

**Synthesis of PMOB material.** PMOB material was prepared by adapting the protocol of Sanchez-Vazquez et al.[38], in which the HCl concentration (Fisher 36 wt %) was decreased to improve the uniformity of the pore channels. Briefly, 3 g of Pluronic P123 triblock copolymer was dissolved in 96 cm$^3$ of water and 1 cm$^3$ of HCl under stirring at 40°C. The BTSB precursor was subsequently added to the surfactant solution, which was stirred at 40°C for a further 72 h. The mixture was then aged at 130°C for 24 h and the resulting solid product filtered, washed three times with deionised water and dried at room temperature.

**Calcination.** PMOB materials were calcined at different calcination temperatures ranging from 350°C and 600°C. The resulting materials are denoted PMOB-X°C, where X represent the calcination temperature. For example, PMOB-400°C was calcined at 400°C for 5 h under static air condition. The sample PMOB-EXT denotes the uncalcined material, after template extraction using several washing with EtOH under reflux condition for 24 h.

**Reaction with API.** 500 mg of PMOB-450°C was mixed with 50 mL of distilled water. After 2 minutes, X drop of pure 1-(3-aminopropyl)imidazole is added to the solution. The solution is stirred for one hour. The color solution changes from yellow to brown color (Figure S7). Then, the material was recovered by filtration and washed with water. The final material was dried at room temperature overnight. The obtained materials are denoted PMOB-450°C-XD.

**Thermogravimetry Analysis.** Thermogravimetric analysis (TGA) was performed using a Stanton Redcroft STA780 thermal analyser on ∼10-20 mg samples under air flow (100 cm$^3$.min$^{-1}$ total flow) during heating at 10 °C.min$^{-1}$ between 20 °C and 1000 °C.

**Porosimetry.** Nitrogen adsorption-desorption was measured using a Quantachrome Nova 2000e porosimeter using NOVAWin software. Samples were degassed at 120 °C for 2 h before analysis by N$_2$ adsorption at −196 °C. Surface areas were calculated from BET model applied over the relative pressure range 0.01–0.2. Pore diameters and volumes were calculated applying the BJH method to the desorption isotherm for relative pressures >0.35.

**Low-Angle XRD.** Low-angle powder XRD patterns were recorded on a PANalytical X'pertPro diffractometer fitted with an X'celerator detector and Cu Kα (1.54 Å) source calibrated against a Si standard (PANalytical). Low angle patterns were recorded for 2θ = 0.3–8° with a step size of 0.01°.

**Transmission Electronic Microscopy.** TEM pictures were obtained with a JEOL 2100 transmission electron microscope operated at 200 kV, with images recorded by a Gatan Ultrascan 1000XP digital camera.

**Conventional NMR**. The solid-state NMR experiments were performed at 9.4 T (i.e. $^1$H Larmor frequency of 400 MHz) on a Bruker BioSpin Avance II NMR spectrometer equipped with 4 or 7 mm double-resonance HX MAS NMR probe for $^{13}$C and $^{29}$Si nuclei, respectively. The rotor was spun at the MAS frequencies, $\nu_R$ = 10 and 5 kHz for $^{13}$C and $^{29}$Si nuclei, respectively. The 1D $^1$H→$^{29}$Si CPMAS spectra of Fig. 4 were recorded using a π/2 $^1$H excitation pulse lasting 4.5 μs and a CP contact time of 5 ms. During the CP transfer, the rf nutation frequency on the $^{29}$Si channel was constant and equal to 35 kHz, whereas the $^1$H nutation frequency was linearly ramped from 27 to 55 kHz. The 1D $^1$H→$^{29}$Si CPMAS spectra result from averaging 2048 transients with a relaxation delay $\tau_{RD}$ = 3 s. The $^{29}$Si isotropic chemical shifts were referenced to tetramethylsilane (TMS). The 1D $^{29}$Si NMR spectra displayed in Fig.4 were simulated using Gaussian lineshapes and TopSpin software in order to determine the signal fractions shown in Figs. S3 and S4 and in Table S2. The 1D $^1$H→$^{13}$C CPMAS spectra of Fig. 5 were recorded using a π/2 $^1$H excitation pulse lasting 3 μs and a CP contact time of 1 ms. During the CP transfer, the rf nutation frequency on the $^{13}$C channel was constant and equal to 50 kHz, whereas the $^1$H nutation frequency was linearly ramped from 41 to 83 kHz. The 1D $^1$H→$^{13}$C CPMAS spectra result from averaging 1024 transients with a relaxation delay $\tau_{RD}$ = 5 s. The $^{13}$C isotropic chemical shifts were referenced to tetramethylsilane (TMS) using the deshielded resonance of adamantane (38.5 ppm) as a secondary reference.

**DNP-NMR.** DNP-NMR experiments were performed on PMOB-450°C and PMOB-450°C-1D samples impregnated with a solution of AMUPol nitroxide biradical.[39] The AMUPol biradical was purchased from SATT Sud-Est (Marseille, France). The samples were prepared by impregnating, during 1 h at room temperature, about 30 mg of PMOB material with 30 μL of 15 mM AMUPol solution in [$^2$H$_6$]-DMSO/H$_2$O (78/22 w/w) mixture. Such mixture forms a glass under temperatures of about 100 K used for DNP-NMR experiments. The formation of such glass insures a uniform distribution of the AMUPol biradicals. For DNP-NMR experiments, the samples were placed in 3.2 mm sapphire rotors since this material is transparent to microwaves at 263 GHz.

The DNP-NMR experiments were performed at 9.4 T on a Bruker BioSpin Avance III DNP-NMR spectrometer equipped with a triple-resonance $^1$H/X/Y 3.2 mm low-temperature MAS probe and a 263 GHz gyrotron.[40] The microwave irradiation was transmitted through a corrugated waveguide to the probe. The microwave power delivered to the sample was ~ 6 W. The samples were spun at MAS frequencies of $\nu_R$ = 8 and 10.5 kHz. The NMR spectra were acquired at a temperature of ~ 105 K, which was stabilized using a Bruker BioSpin MAS cooling system.

The build-up curves of the $^1$H polarization with and without microwave irradiation were measured using a saturation-recovery experiment, involving a burst of presaturation pulses (50 π/2 pulses separated by 1.0 ms) followed by a relaxation delay (ranging from 0.1 to 60 ms). For each relaxation delay, 4 transients were accumulated. The signal intensity of the saturation-recovery experiments as function the relaxation delay was fitted to stretched-exponential function.[41,42] Such fit yielded the effective build-up time of the DNP-enhanced $^1$H polarization, $T_{B,on}(^1H)$, for the experiment with microwave irradiation and the $^1$H longitudinal relaxation time, $T_{B,off}(^1H)$, for that without microwave irradiation. The 1D $^1$H→$^{13}$C CPMAS spectra were acquired with and without microwave irradiation. SPINAL-64 $^1$H decoupling[43] with a radiofrequency (rf) nutation frequency $\nu_{1,dec}(^1H)$ = 90 kHz was applied during the acquisition. The $^1$H π/2 pulse duration and the CP contact time were 2.8 µs and 1.5 ms, respectively. During the CP transfer, the $^1$H rf nutation frequency was linearly ramped from 66 to 93 kHz, whereas the rf nutation frequency on $^{13}$C channel was constant and equal to 76 kHz. For $^1$H→$^{13}$C CPMAS experiments with and without microwave irradiation, the relaxation delay, $\tau_{RD}$, was fixed to 1.3 times $T_{B,on}(^1H)$ and $T_{B,off}(^1H)$, respectively.[44] The $^{13}$C isotropic chemical shifts were referenced to TMS using the deshielded resonance of adamantane (38.5 ppm) as a secondary reference.

**DRIFTS.** The IR measurements were recorded with a Nicolet Protege System 460 equipped with a MCT detector in transmission mode. The analysis was made with in-situ cell, under He flow at 200°C, to remove the physisorbed water present in the material. All samples were acquired after 2 h under He flow. All spectra were normalized with respect to the band at 1850 cm$^{-1}$ (characteristic vibration of SBA-15) for better visualization of changing in silanol region.

**XPS**. XPS characterization was carried out with a Kratos AXIS UltraDLD instrument using a monochromatic Al KaX-ray source (10 mA,12 kV). High resolution spectra were obtained using a 40 eV pass energy. All spectra were charge corrected to give to the adventitious C 1s component a binding energy of 284.8 eV. Spectra were analyzed using Casa XPSsoftware (version 2.3.16, Casa Software Ltd.). Quantifications were performed after a Shirley background subtraction.

### 3. Results and discussion

#### 3.1. Structural characterization of calcined PMOB

**TGA**. We first characterize the structural modifications of PMOB during calcination. TGA analysis (under air) of as-synthesized PMOB material show a degradation of the template, i.e. the Pluronic P123 triblock copolymer, around 300°C, which is followed by the oxidation of the benzene rings around 600°C (Figure S1). Therefore, we focus here on the PMOB materials calcined at temperature ranging from 350 to 600°C. The PMOB sample prepared by calcination at temperature *T* is denoted PMOB-*T*, whereas PMOB-EXT denotes the uncalcined PMOB sample, for which the template was extracted under ethanol reflux.

**Porosimetry.** As seen in Figure 1, all of the PMOB materials keep the type IV isotherms with H1 hysteresis loops, according to the IUPAC classification. Such results indicate that all these materials contain ordered mesopores. The isotherms of PMOB-EXT, PMOB-350°C and PMOB-400°C materials are similar to SBA-15 isotherm, confirming the presence of long channel with regular large pore diameter. When the calcination temperature increase above 450°C, the isotherm profile change slightly to an isotherm corresponding to ordered porous network with interconnections between channels and large pore size (as in KIT-6 material). These interconnections stem from the partial degradation of the benzene-silica wall above 450°C. As seen in Table S1, the specific surface area and the porous volume decrease for increasing calcination temperature. Furthermore, as seen in Figure S2 and Table S1, the average pore diameter decreases from 4.0 to 3.6 nm, when the calcination temperature increases from 350 to 600°C.

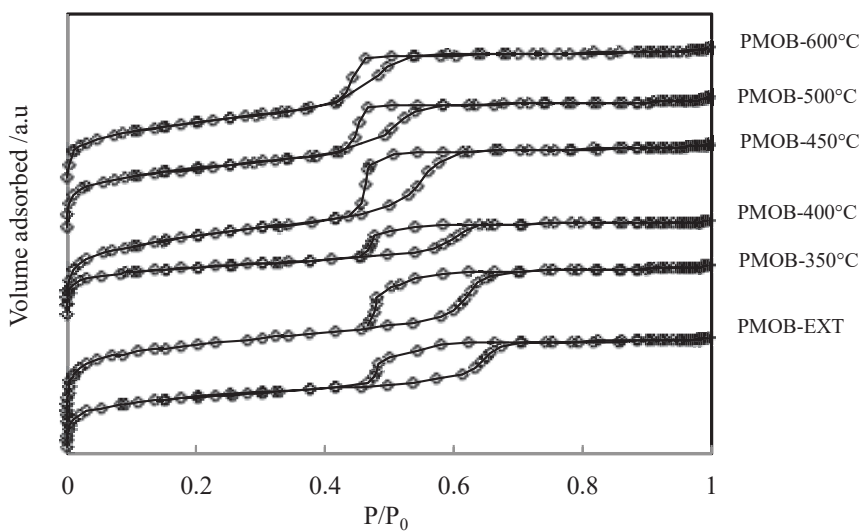

**Figure 1.** $N_2$ adsorption-desorption isotherms of PMOB-EXT and PMOB-$T$ materials with $T$ = 350, 400, 450, 500 and 600°C. The isotherms are vertically separated for better visualization.

**XRD and TEM**. The powder XRD pattern recorded at low-angle values for the different PMOB materials are displayed in Figure 2. The low angle diffractions, corresponding to 100, 110 and 200 planar symmetries of the p6mm structure, show a good degree of ordering in the mesostructures, even for samples calcined at high temperature. The $d_{100}$ spacing recorded for the different materials shift at higher angle when calcination temperature increases (see the inset A of Figure 2). This is consistent with a decrease of the unit cell parameter and corresponding pore diameter (see Table S1). TEM pictures (Figure 3) show the porous structure of the PMOB-EXT material, which is preserved after calcination at 450°C.

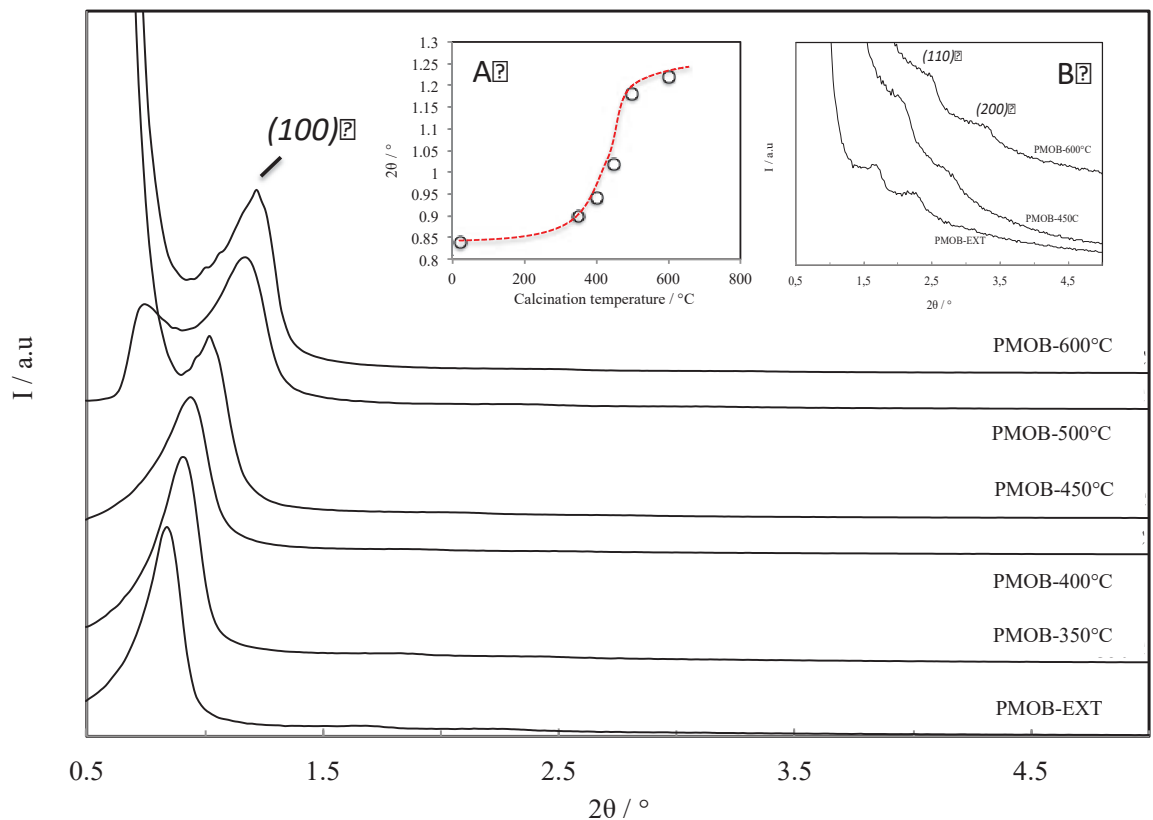

**Figure 2.** Powder XRD patterns of PMOB materials. The inset A shows the variation of 2θ angle corresponding to the planar symmetrie (100) as function of calcination temperatutre. The inset B shows a zoom of the XRD patterns on the angle region corresponding to the planar symmetrie (110) and (200) for PMOB-EXT, PMOB-450°C and PMOB-650°C.

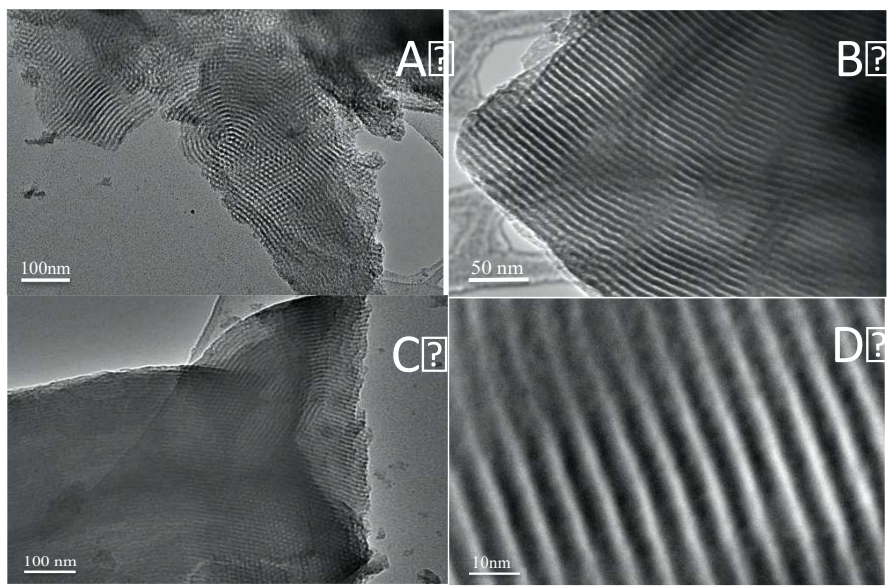

**Figure 3.** TEM pictures of (A,C) PMOB-EXT and (B,D) PMOB-450°C.

**Conventional NMR.** Figure 4 displays the $^1$H→$^{29}$Si CPMAS NMR spectra of PMOB materials under magic-angle spinning (MAS). The $^{29}$Si spectrum of PMOB-EXT displays intense signals resonating at −61, −70 and −78 ppm, which are ascribed to silicon sites T$^1$ [(SiO)SiPh(OH)$_2$], T$^2$ [SiPh(OH)(OSi)$_2$] and T$^3$ [SiPh(OSi)$_3$], respectively.[12] That sample also displays weak $^{29}$Si signals at −91 and −100 ppm, which are assigned to Q$^2$ [(SiO)$_2$Si(OH)$_2$] and Q$^3$ [(SiO)$_3$SiOH] environments.[45] Even if the signal intensity in $^1$H→$^{29}$Si CPMAS NMR spectra depends on the CP transfer efficiency between $^1$H and $^{29}$Si nuclei and hence, these spectra are only semi-quantitative, the weak intensity of Q sites with respect to T ones indicates that in PMO-EXT material, most Si atoms are covalently linked to benzene ring. During the calcination, when the temperature increases, the amount of *T* sites decreases, whereas that of Q sites increases (see Figures S3, S4 and Table S2). These changes indicate the oxidation of T sites into Q ones. According to $^{29}$Si NMR data, the fraction of T sites in PMOB-500°C and PMOB-600°C materials is close to zero, which indicates the cleavage of almost all Si-C bonds.

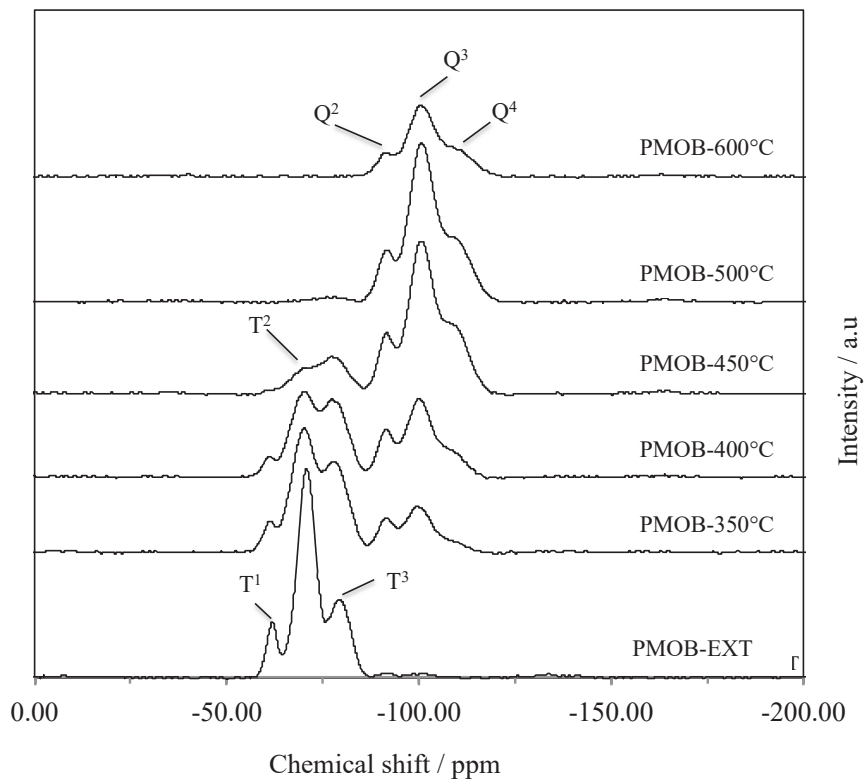

**Figure 4.** Conventional 1D $^1H\rightarrow^{29}Si$ CPMAS NMR spectra of PMOB materials at $B_0$ = 9.4 T with $\nu_R$ = 5 kHz.

The $^1H\rightarrow^{13}C$ CPMAS spectra of PMOB materials are shown in Figure 5. The spectrum of PMOB-EXT is dominated by a peak at 134 ppm, which is assigned to the CH site of the phenylene ring.[7,14] The weak shoulder at 129 ppm is assigned to the quaternary C site of the phenylene ring. This signal is less intense than that of CH since (i) the number of quaternary C atoms in phenylene ring is twice lower than that of CH groups and (ii) the $^1H\rightarrow^{13}C$ CPMAS transfer is less efficient for quaternary C atoms than for CH groups as the quaternary $^{13}C$ nuclei are more distant from protons than the CH ones. After several extractions with ethanolic HCl solution, PMOB-EXT still exhibit signals at 70, 59.7 and 17.5 ppm. The signals at 59.7 and 17.5 ppm are assigned to the $CH_2$ and $CH_3$ sites of ethoxy groups, whereas those at 70 and 18 ppm are assigned to the $CH_2$ and $CH_3$ sites of template Pluronic 123.[46,47]

After calcination, the $^{13}$C signals of ethoxy groups and template are not visible, which indicates that the calcination evacuates those species from the PMOB pores. Besides, the relative intensity of the peak at 129 ppm increases with respect to that at 134 ppm at higher calcination temperature. Such increase stems from the cleavage of Si-C bonds and the formation of Si-Ph groups.[47] Such cleavage of the Si-C bonds during calcination is consistent with the decrease of the integrated intensity of T signals in the $^1H\rightarrow{}^{29}Si$ CPMAS spectrum at higher calcination temperature (see Figure S3). Furthermore, new peaks at 154, 121 and 114 ppm appear during the calcination and their relative intensity with respect to the signal at 134 ppm increases for increasing calcination temperature. These peaks are assigned to phenolic groups, such as Si-PhOH or Si-Ph(OH)-Si, or polyphenolic groups, such as Si-Ph(OH)$_n$ or Si-Ph(OH)$_n$-Si, resulting from the oxidation of the phenylene bridge, with or without the cleavage of the Si-C bonds.[47,48] We can also notice a broadening of the peak at 134 ppm for higher calcination temperature. Such broadening indicates a greater distribution of local environments for the aromatic CH sites at higher calcination temperature. Such increased disorder is consistent with the cleavage of some C-Si bonds and the formation of phenol groups during calcination. No $^{13}$C carbonyl signal, which usually resonate in the range 170-220 ppm, is detected in the conventional $^1H\rightarrow{}^{13}C$ CPMAS spectra of PMOB material. The $^1H\rightarrow{}^{13}C$ CPMAS spectrum of PMOB-600°C does not exhibit any $^{13}$C signal in the aromatic region. Such result is consistent with the lack of T signals in the $^{29}$Si spectra of Figure S3 and indicates that most C atoms are eliminated in the form of volatile molecules ($CO_2$, etc) during the calcination at 600°C.

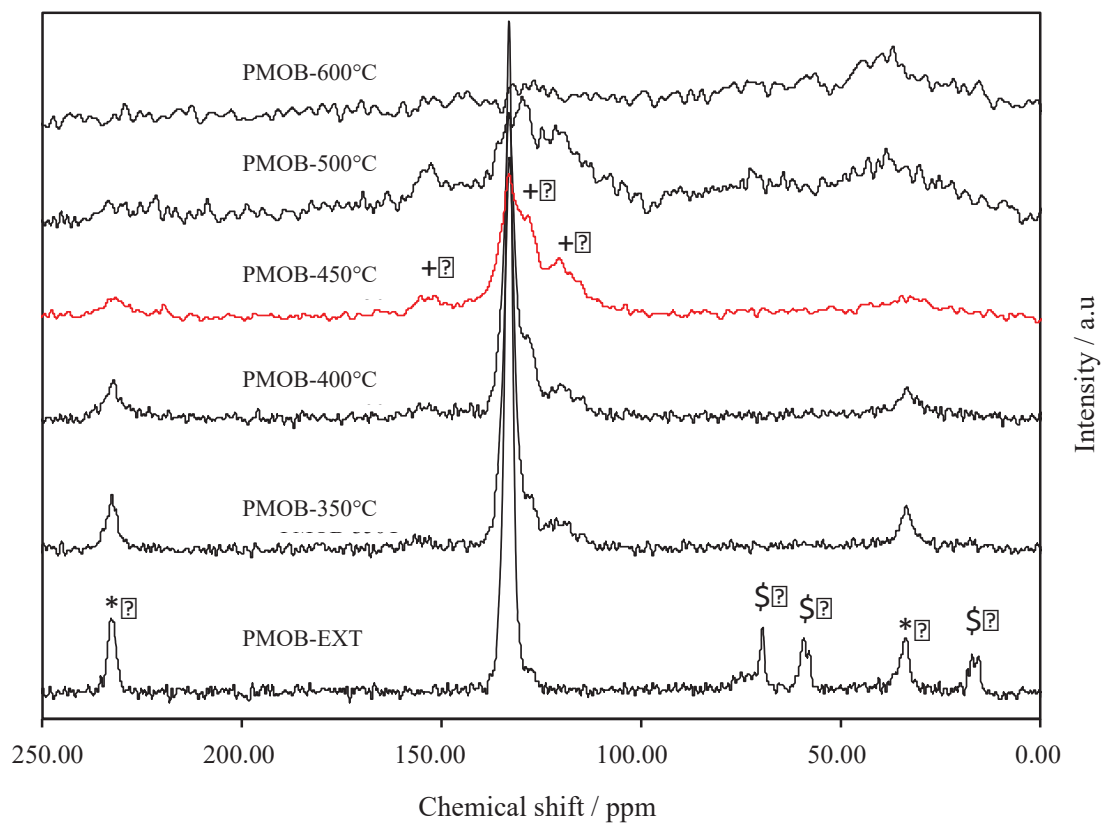

**Figure 5.** 1D $^1H \rightarrow {}^{13}C$ CPMAS NMR of PMOB materials. The * symbol denotes the spinning sidebands of aromatic group at 138 ppm. The $ symbol denotes the residual organic template or ethoxy groups. The + symbol denotes new signals observed after calcination, which are notably intense in the spectrum of PMOB-450°C.

**DNP-NMR.** In order to identify more dilute chemical functionalities formed during the calcination of PMOB material, DNP-enhanced $^1H \rightarrow {}^{13}C$ CPMAS experiments were conducted on PMOB-450°C. The sample was impregnated with a solution of nitroxide biradical, AMUPol, which served as DNP polarizing agent.[39] The build-up of the $^1H$ polarization is faster with microwave irradiation than without it (see the inset in Figure 6a). Furthermore, as seen in Figure 6, microwave irradiation yields a 60-fold signal enhancement for dimethylsulfoxide (DMSO) solvent used for the impregnation in $^1H \rightarrow {}^{13}C$ CPMAS experiment but only a 27-fold one for the aromatic $^{13}$C sites of PMOB for the employed relaxation delays (see the experimental section). These observations indicate a heterogeneous polarization in the sample. The spin diffusion transports the DNP-enhanced $^1$H polarization but the size of domains to be polarized exceeds the spin diffusion length.[33,42] The AMUPol radical does not thoroughly penetrate into the pores. As the dimensions (about 1.28 nm for the O−O distance) of the AMUPol molecules are smaller than the pore size of PMOB-450°C (3.6 nm, see Table S1), the inhomogeneous distribution of radicals within the pores must stem from the occlusion of some pores during the calcination. These occlusions are consistent with the 44% reduction of the specific surface area of PMOB material, when the calcination temperature increases from 350°C to 450°C. Note that the DNP-enhanced $^1$H→$^{13}$C CPMAS spectrum was acquired with a relaxation delay shorter than that used without microwave irradiation owing to faster build-up of the $^1$H polarization with microwave irradiation. Higher signal enhancement will be measured for identical relaxation delay with and without microwave irradiation.

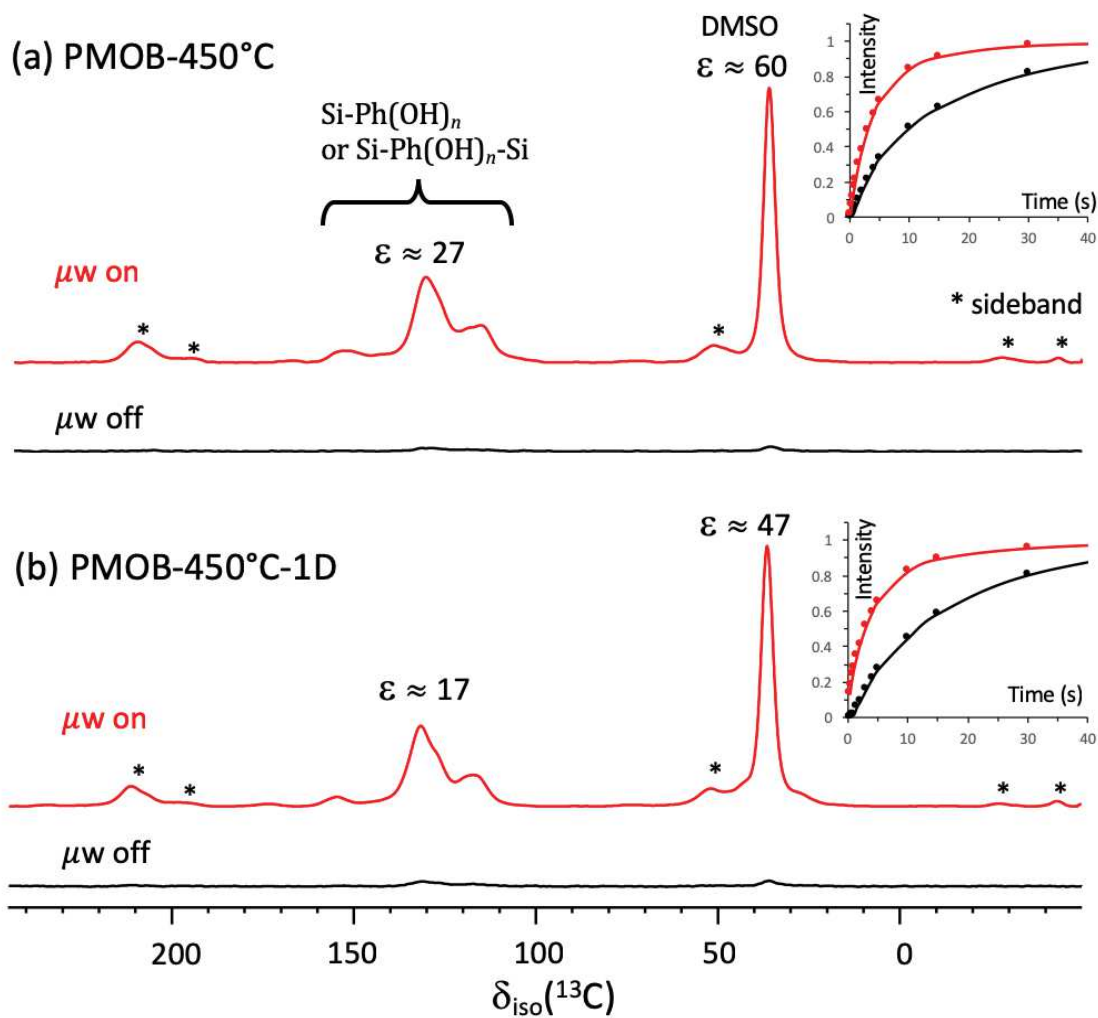

**Figure 6.** 1D $^1$H→$^{13}$C CPMAS spectra of (a) PMOB-450°C and (b) PMOB-450°C-1D impregnated with 15 mM AMUPol solution in [$^2$H$_6$]-DMSO/H$_2$O mixture at 9.4 T and 105 K with a MAS frequency of 8 kHz. The spectra acquired with and without µw irradiation are shown by red and black lines, respectively. For both samples, the build-up curves of the $^1$H signal as function of the polarization delay with (red) and without (black) µw irradiation are displayed as inset. The build-up curves are normalized so that the asymptotic intensity is equal to unity. The $\tau_{RD}$ delays were equal to 11.5 and 19.1 s with and without microwave irradiation in subfigure a and 7.4 and 21.6 s with and without microwave irradiation in subfigure b. The number of transients were equal to 128 and 512 in subfigures a and b, respectively, in the case of microwave irradiation, and 64 in the absence of microwave irradiation. The intensity of the spectra are normalized with respect to the number of transients.

The sensitivity enhancement provided by DNP allows the detection of additional $^{13}$C signals (see Figures 6 and 7) compared to conventional NMR. The signals ranging from 55 to 80 ppm are assigned to aliphatic and allylic alcohol functionalities. The peak at 169 ppm is assigned to carboxylic acids, in which the C=O double bonds is conjugated with C=C double bonds. These additional species are formed during the oxidative cleavage of benzene ring. At a MAS frequency of 8 kHz, the first-order spinning sideband of the $^{13}$C aromatic signal extends from 190 and 220 ppm, which can mask the signals of aldehyde and ketone groups. Hence, we also acquired the DNP-enhanced $^1$H→$^{13}$C CPMAS spectrum of PMOB-450°C at MAS frequency of 10.5 kHz (see Figure 7). Such spectrum exhibits an additional weak peak extending from 190 to 200 ppm, which is ascribed to aldehyde and ketone groups. However, the intensity of the peak of ketone, aldehyde and alcohol groups are much weaker than that of carboxylic acid. Thus, the oxidative cleavage of benzene ring produces mostly carboxylic acid functionalities.

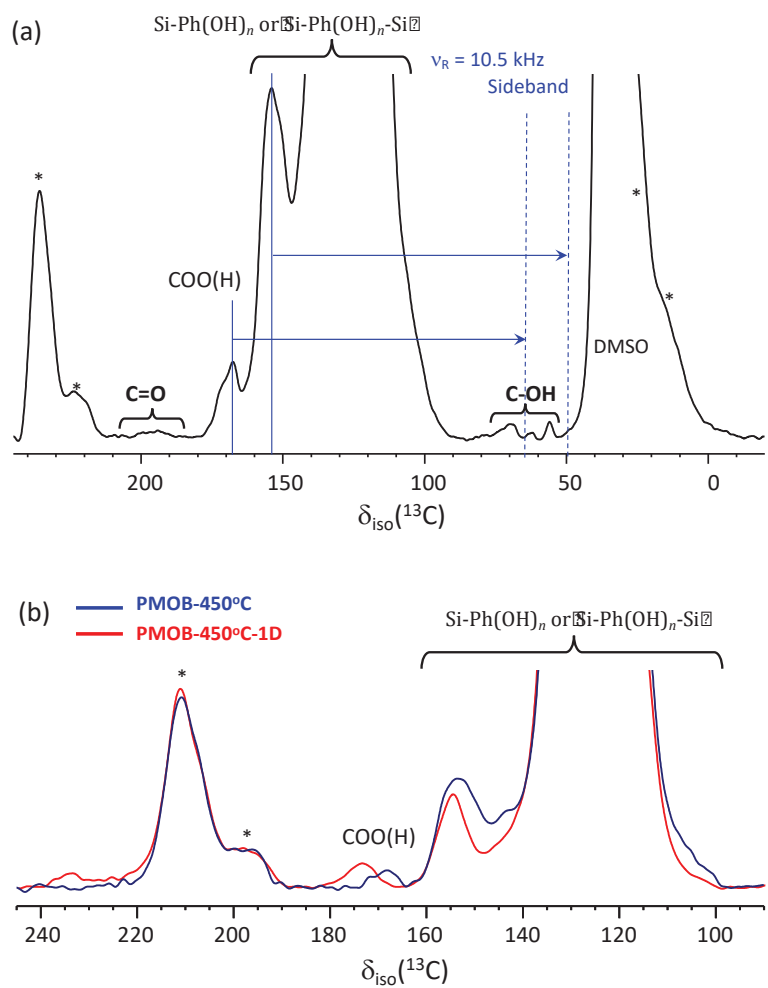

**Figure 7.** (a) Vertical expansion of the DNP-enhanced 1D $^1$H→$^{13}$C CPMAS spectra of PMOB-450°C impregnated with 15 mM AMUPol solution in [$^2$H$_6$]-DMSO/H$_2$O mixture at 9.4 T and 105 K with a MAS frequency of 10.5 kHz and µw irradiation. The other experimental parameters are identical to those of Fig. 6. (b) Expansion of 240-90 ppm region of the DNP-enhanced 1D $^1$H→$^{13}$C CPMAS spectra of PMOB-450°C and PMOB-450°C-1D impregnated with 15 mM AMUPol solution in [$^2$H$_6$]-DMSO/H$_2$O mixture at 9.4 T and 105 K with a MAS frequency of 8 kHz. The spectra are displayed with the same intensity for the peak at 134 ppm. The other experimental parameters are given the caption of Figure 6 and the experimental section.

The nature of organic entities formed during the calcination step was also investigated using DRIFTS. The analysis was made with in-situ cell, under He flow at 200°C, to remove the physisorbed water present in the material. The DRIFTS spectra are shown in Figure 8. The spectrum of PMOB-EXT displays a broad and intense band between 3740 and 3200 cm$^{-1}$ assigned to the stretching mode of the various OH groups.[49,50] Bands at 3060 and 3015 cm$^{-1}$ are assigned to the stretching modes of C-H bonds in benzene rings, whereas those at 2964, 2934 and 2870 cm$^{-1}$ are assigned to the stretching modes of C-H bonds in the template and non-hydrolysed ethoxy groups.[47,51] The band at 1384 cm$^{-1}$ corresponds the vibration of the covalent bonds between C atoms in disilylbenzene ring.[49] Finally the intense band starting below 1300 cm$^{-1}$ is assigned to the vibration of Si-O bonds.

From calcination at 350°C of PMOB material, the bands assigned to the template strongly decreases, which indicates its decomposition and from 400°C, we observe a total decomposition of the template. Conversely a band at 3650 cm$^{-1}$ assigned to the stretching of O-H bonds in phenol moieties appears. Such result is consistent with the assignment of $^1$H→$^{13}$C CPMAS spectra (see Figure 5). The calcination also produces a band at 1710 cm$^{-1}$, which is assigned to the stretching of C=O bond conjugated with C=C bond in α,β-unsaturated carboxylic acid. The intensity of this band is maximal for calcination under air at 450°C. Hence, DRIFTS data confirms the DNP-NMR data. For the calcined materials, four bands are also detected at 1595, 1570, 1506 and 1477 cm$^{-1}$. They are assigned to the stretching modes of carbon-carbon bonds in aromatic rings bonded to hydroxyl and silyl groups as well as alkenes.

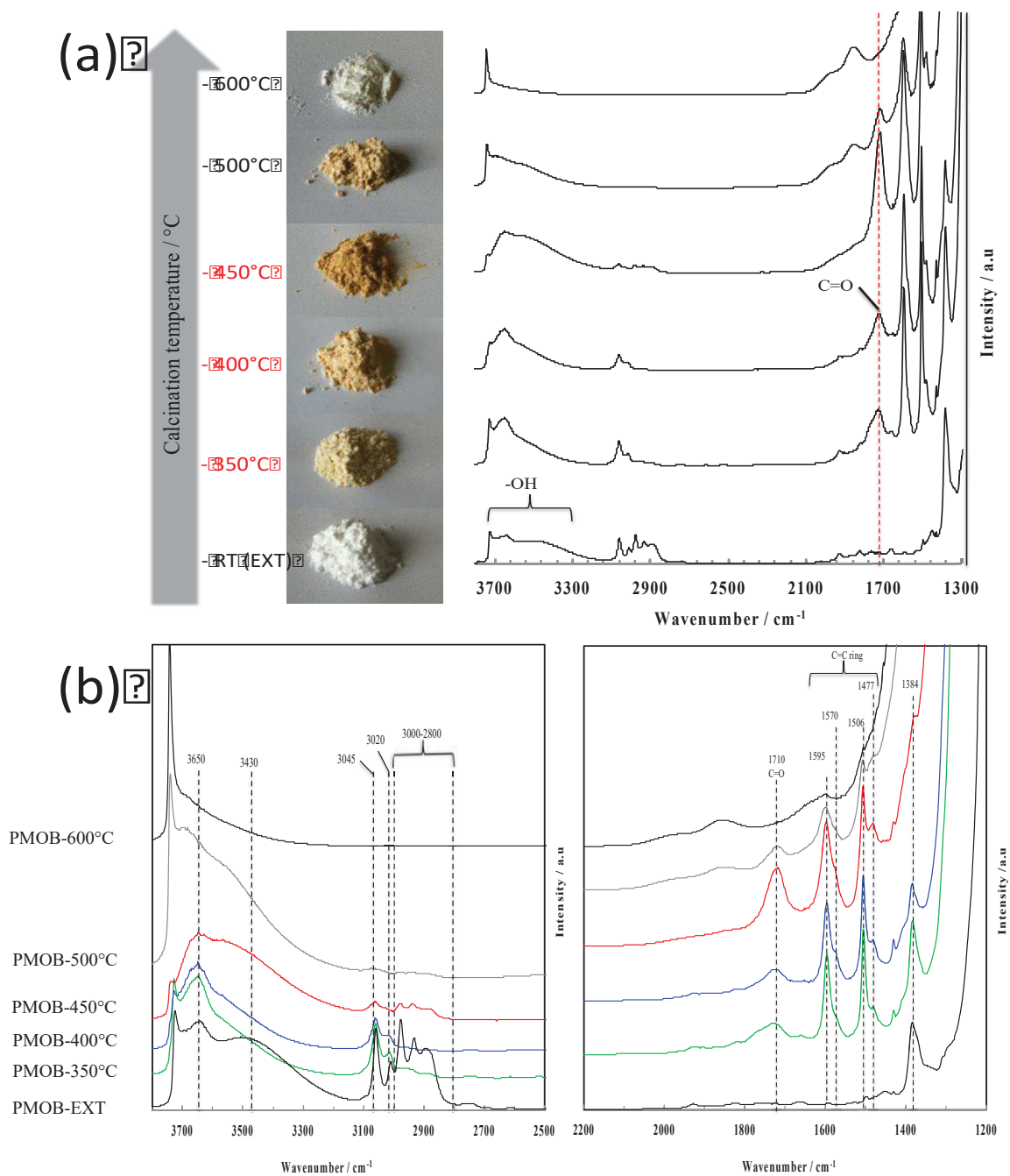

**Figure 8.** (a) DRIFTS spectra of PMOB materials. (b) Expansion of 3800-2500 cm$^{-1}$ and 2200-1200 cm$^{-1}$ regions of the DRIFTS spectra shown in subfigure a.

**XPS**. As seen in Figure S5, the intensity of C1s XPS spectra decreases for calcination temperature above 450°C. Such observation is consistent with the decreased $^1H\rightarrow{}^{29}Si$ CPMAS signals of T sites (see Figures 4 and S3) during calcination as well as the decrease of the $^1H\rightarrow{}^{13}C$ CPMAS signal of phenylene ring at high calcination temperature (see Figure 5). The calcination at 350°C decreases the intensity of the peaks assigned to C−O and C−OH (see Figures S5 and S6). This decrease confirms the elimination of ethoxy groups and template molecules. The peak of C−O groups increases again at higher calcination temperature because of the formation of (poly)phenolic groups (see Fig. S6). Furthermore, PMOB-450°C exhibits an additional peak at 289.6 eV, which is assigned to carboxylate group. This observation is consistent with the $^1H\rightarrow{}^{13}C$ CPMAS spectra of Fig. 7 and confirms the formation of carboxylic acid groups by oxidative cleavage of benzene rings.

### 3.2. Reaction with API

**Optical absorbance**. We investigated the reaction of PMOB-450°C with API in distilled water at room temperature. After the addition of a single drop of API to water solution containing 500 mg of PMOB-450°C, the yellow powder become brown (Figure S7). Such color change may result from the partial deprotonation of phenol by the $NH_2$ group of API since (i) the protonation state of the phenol-based PMOs affect their optical absorbance and (ii) the $pK_a$ constants of phenol and primary amines are 9.95 and 10.66, respectively.[52]

**Porosimetry.** After recovering the solid, the $N_2$ adsorption-desorption measurements shown in Figure S8 indicate that the adsorbed $N_2$ volume decreases with the number of API drops added to PMOB-450°C. After the addition of 4 API drops, the specific surface area and the total pore volume of PMOB-450°C materials are strongly reduced (see Table S3). The addition of four API drops reduces the specific surface area and the total pore volume by a factor larger than 10. Hence, the pores of PMOB-450°C-4D seems to be almost filled with API.

**DNP-NMR.** Like for PMOB-450°C, we acquired DNP-enhanced 1D $^1$H→$^{13}$C CPMAS spectrum of PMOB-450°C-1D impregnated with AMUPol solution. Here again, we measured a build-up of $^1$H polarization, which is faster with microwave irradiation than without and a larger signal enhancement for the DMSO ($\varepsilon \approx 47$) than for the aromatic carbons ($\varepsilon \approx 17$) (see Figure 6). These observations indicate again a heterogeneous $^1$H polarization in the sample, which result from the occlusion of some pores of PMOB-450°C-1D, in which AMUPol radicals cannot penetrate. Furthermore, the dimensions of the regions inaccessible to these polarizing agents exceed the distance, over which $^1$H-$^1$H spin diffusion can transport the DNP-enhanced polarization. Nevertheless, PMOB-450°C-1D exhibits a faster build-up of $^1$H polarization than PMOB-450°C. This difference can stem from the presence of protonated API and water in the pores, which accelerates the $^1$H-$^1$H spin diffusion.

The DNP-enhanced 1D $^1$H→$^{13}$C CPMAS spectrum of PMOB-450°C-1D is almost identical to that of PMOB-450°C (compare Figures 6a and b). The signals of [$^2$H$_6$]-DMSO and the first-order spinning sideband of the aromatic carbons overlap with those of propyl chain of API resonating at 34, 38 and 44 ppm in CDCl$_3$ solution. Similarly the signals of the aromatic carbon nuclei of PMOB-450°C overlap those of imidazole ring resonating at 119, 129 and 137 ppm in solution. As seen in Figure 7b, after the addition of API, the signal of $^{13}$C sites bonded to oxygen in silylphenol moieties is deshielded and its relative intensity with respect to the signal at 134 ppm decreases. Such modification stems from the partial deprotonation of phenol by the NH$_2$ group of API, in agreement with the color change shown in Figure 7. This deprotonation reduces the efficiency of the $^1$H→$^{13}$C polarization transfer for the $^{13}$C atom bonded to oxygen, which explains the decrease in its relative intensity. Furthermore, after the addition of API, the carbonyl signal is shifted towards higher isotropic chemical shift. Such shift can stem from (i) the deprotonation of the carboxylic acid or (ii) the hydroamination of the C=C bond conjugated with C=O bond of carboxylic acid. Given the $pK_a$ of acrylic acid is 4.25, the primary amine of API deprotonates the unsaturated carboxylic acids of PMOB-450°C. Furthermore, we do not observe peak around 163 ppm corresponding to the C=O function of an amide conjugated to a C=C bond and there is a decrease in the intensity near 150 ppm, the isotropic chemical shift of imine function conjugated to a C=C bond. Hence, no formation of amide or imine groups by reaction between carboxylic groups of PMOB-450°C and API is detected using DNP-enhanced 1D $^1$H→$^{13}$C CPMAS experiments.

**DRIFTS.** Figure S9 shows the DRIFTS spectra of the PMOB-450°C, -1D and -4D materials. After the addition of API, two bands are detected at 3338 and 3275 cm$^{-1}$, which are ascribed to the asymmetric and symmetric stretching mode of the NH$_2$ group of API. Furthermore, we also detect a band at 3105 cm$^{-1}$ assigned to the stretching vibration of C-H bond in imidazole ring and bands at 2921 and 2852 cm$^{-1}$ assigned to the stretching modes of the C-H bond in the propyl chain of API. Hence, the DRFITS data confirms the presence of API in the pores of PMOB-450°C-1D and -4D. After the addition of API, we observe a decreased intensity for the peak resonating at 1708 cm$^{-1}$ assigned to the stretching mode of C=O bond of carboxylic acid and the appearance of peaks at 1640 and 1430 cm$^{-1}$ assigned to the antisymmetric and symmetric modes of the carboxylate group.[53,54] These latter peaks are more intense for PMOB-450°C-4D than for PMOB-450°C-1D. These observations confirm the deprotonation of carboxylic groups by reaction with API.

**XPS.** The N 1s XPS spectrum (shown in Fig. S10) of PMOB-450°C-1D exhibits peaks at 402 eV and 399.5 eV ascribed to -C-N- and -C=N- groups of imidazole moiety.[55–59] The peak at 400.7 eV is assigned to the amine group of API .[60,61] Hence, the XPS spectrum confirms the presence of API molecules in the pores of PMOB-450°C-1D.

## 4. Conclusions

We characterized the structure of calcined PMOB using various techniques, including solid-state (DNP)-NMR, DRIFTS, XRD, XPS, TGA and porosimetry. Calcination leads to the removal of non-hydrolyzed ethoxy groups and template molecules. These entities are already completely removed at 350°C. The calcination also leads to the cleavage of Si-C bonds and the oxidation of T sites into Q ones. The T sites disappear above 500°C. The calcination also oxidizes the phenylene bridges into (poly)phenolic groups. The amount of these groups is maximal at 450°C. DNP-enhanced 1D $^1H \rightarrow ^{13}C$ CPMAS spectra also revealed the oxidative cleavage of some benzene rings at 450°C leading to the formation of carboxylic acids as well as allylic alcohols, ketones and aldehydes in much lesser extent. The formation of carboxylic functionalities during calcination is confirmed by DRIFTS and XPS data. The degradation of the benzene rings produces interconnections between the pores. Although the surface area and the pore volume of PMOB decreases for increasing calcination temperatures above 350°C, the PMOB materials still exhibit porous structure after calcination at 600°C. DNP-NMR and DRIFTS experiments also reveal that API amine groups deprotonate the phenol and carboxylic acid functionalities of PMOB-450°C in water at room temperature. Nevertheless, no formation of covalent bonds between API and the calcined PMOB was detected in these experimental conditions.

## ASSOCIATED CONTENT

The Supporting information is available free of charge on the ACS Publications website.

TGA; surface areas, pore volumes, pore diameters, unit cell parameters; $^{29}$Si signal fractions of Q and T sites; XPS data; color change after the addition of API; porosimetry and DRIFT spectra of PMOB-450°C after the addition of API.

## AUTHOR INFORMATION

**Corresponding Author**

*Email: cyril.pirez@gmail.com (C. P.)

*Email: olivier.lafon@univ-lille.fr (O. L.)

**ORCID**

Hiroki Nagashima: 0000-0002-7042-9051

Franck Dumeignil: 0000-0001-9727-8196

Olivier Lafon: 0000-0002-5214-4060

## ACKNOWLEDGMENT

Chevreul Institute (FR 2638), Ministère de l'Enseignement Supérieur, de la Recherche et de l'Innovation, Hauts-de-France Region and FEDER are acknowledged for supporting and funding partially this work. Financial support from the IR-RMN-THC FR-3050 CNRS for conducting the research is gratefully acknowledged. Authors also thank contracts ANR-17-ERC2-0022 (EOS) and ANR-18-CE08-0015-01 (ThinGlass). This project has received funding from the European

Union's Horizon 2020 research and innovation program under grant agreement No 731019 (EUSMI). OL acknowledge financial support from Institut Universitaire de France (IUF).

REFERENCES

(1) Nothling, M. D.; Ganesan, A.; Condic-Jurkic, K.; Pressly, E.; Davalos, A.; Gotrik, M. R.; Xiao, Z.; Khoshdel, E.; Hawker, C. J.; O'Mara, M. L.; et al. Simple Design of an Enzyme-Inspired Supported Catalyst Based on a Catalytic Triad. *Chem* **2017**, *2* (5), 732–745. https://doi.org/10.1016/j.chempr.2017.04.004.

(2) Melde, B. J.; Holland, B. T.; Blanford, C. F.; Stein, A. Mesoporous Sieves with Unified Hybrid Inorganic/Organic Frameworks. *Chem. Mater.* **1999**, *11* (11), 3302–3308. https://doi.org/10.1021/cm9903935.

(3) Asefa, T.; MacLachlan, M. J.; Coombs, N.; Ozin, G. A. Periodic Mesoporous Organosilicas with Organic Groups inside the Channel Walls. *Nature* **1999**, *402* (6764), 867–871. https://doi.org/10.1038/47229.

(4) Inagaki, S.; Guan, S.; Fukushima, Y.; Ohsuna, T.; Terasaki, O. Novel Mesoporous Materials with a Uniform Distribution of Organic Groups and Inorganic Oxide in Their Frameworks. *J. Am. Chem. Soc.* **1999**, *121* (41), 9611–9614. https://doi.org/10.1021/ja9916658.

(5) Asefa, T.; MacLachlan, M. J.; Grondey, H.; Coombs, N.; Ozin, G. A. Metamorphic Channels in Periodic Mesoporous Methylenesilica. *Angew. Chem. Int. Ed.* **2000**, *39* (10), 1808–1811. https://doi.org/10.1002/(SICI)1521-3773(20000515)39:10<1808::AID-ANIE1808>3.0.CO;2-G.

(6) Guan, S.; Inagaki, S.; Ohsuna, T.; Terasaki, O. Cubic Hybrid Organic−Inorganic Mesoporous Crystal with a Decaoctahedral Shape. *J. Am. Chem. Soc.* **2000**, *122* (23), 5660–5661. https://doi.org/10.1021/ja000839e.

(7) Burleigh, M. C.; Markowitz, M. A.; Jayasundera, S.; Spector, M. S.; Thomas, C. W.; Gaber, B. P. Mechanical and Hydrothermal Stabilities of Aged Periodic Mesoporous Organosilicas. *J. Phys. Chem. B* **2003**, *107* (46), 12628–12634. https://doi.org/10.1021/jp035189q.

(8) Cho, E.-B.; Char, K. Macromolecular Templating Approach for the Synthesis of Hydrothermally Stable Mesoporous Organosilicas with High Periodicity and Thick Framework Walls. *Chem. Mater.* **2004**, *16* (2), 270–275. https://doi.org/10.1021/cm0346733.

(9) Nakajima, K.; Lu, D.; Kondo, J. N.; Tomita, I.; Inagaki, S.; Hara, M.; Hayashi, S.; Domen, K. Synthesis of Highly Ordered Hybrid Mesoporous Material Containing Etenylene (–CH=CH–) within the Silicate Framework. *Chem. Lett.* **2003**, *32* (10), 950–951. https://doi.org/10.1246/cl.2003.950.

(10) Wang, W.; Xie, S.; Zhou, W.; Sayari, A. Synthesis of Periodic Mesoporous Ethylenesilica under Acidic Conditions. *Chem. Mater.* **2004**, *16* (9), 1756–1762. https://doi.org/10.1021/cm035235z.

(11) Burleigh, M. C.; Jayasundera, S.; Thomas, C. W.; Spector, M. S.; Markowitz, M. A.; Gaber, B. P. A Versatile Synthetic Approach to Periodic Mesoporous Organosilicas. *Colloid Polym. Sci.* **2004**, *282* (7), 728–733. https://doi.org/10.1007/s00396-003-1004-0.

(12) Yoshina-Ishii, C.; Asefa, T.; Coombs, N.; MacLachlan, M. J.; Ozin, G. A. Periodic Mesoporous Organosilicas, PMOs: Fusion of Organic and Inorganic Chemistry 'inside' the Channel Walls of Hexagonal Mesoporous Silica. *Chem. Commun.* **1999**, No. 24, 2539–2540. https://doi.org/10.1039/A908252B.

(13) Inagaki, S.; Guan, S.; Ohsuna, T.; Terasaki, O. An Ordered Mesoporous Organosilica Hybrid Material with a Crystal-like Wall Structure. *Nature* **2002**, *416* (6878), 304–307. https://doi.org/10.1038/416304a.

(14) Goto, Y.; Inagaki, S. Synthesis of Large-Pore Phenylene-Bridged Mesoporous Organosilica Using Triblock Copolymer Surfactant. *Chem. Commun.* **2002**, No. 20, 2410–2411. https://doi.org/10.1039/B207825B.

(15) Yang, Q.; Kapoor, M. P.; Inagaki, S. Sulfuric Acid-Functionalized Mesoporous Benzene−Silica with a Molecular-Scale Periodicity in the Walls. *J. Am. Chem. Soc.* **2002**, *124* (33), 9694–9695. https://doi.org/10.1021/ja026799r.

(16) Kapoor, M. P.; Inagaki, S.; Ikeda, S.; Kakiuchi, K.; Suda, M.; Shimada, T. An Alternate Route for the Synthesis of Hybrid Mesoporous Organosilica with Crystal-Like Pore Walls from Allylorganosilane Precursors. *J. Am. Chem. Soc.* **2005**, *127* (22), 8174–8178. https://doi.org/10.1021/ja043062o.

(17) Sayari, A.; Wang, W. Molecularly Ordered Nanoporous Organosilicates Prepared with and without Surfactants. *J. Am. Chem. Soc.* **2005**, *127* (35), 12194–12195. https://doi.org/10.1021/ja054103z.

(18) Kapoor, M. P.; Yang, Q.; Inagaki, S. Self-Assembly of Biphenylene-Bridged Hybrid Mesoporous Solid with Molecular-Scale Periodicity in the Pore Walls. *J. Am. Chem. Soc.* **2002**, *124* (51), 15176–15177. https://doi.org/10.1021/ja0290678.

(19) Okamoto, K.; Goto, Y.; Inagaki, S. Self-Organization of Crystal-like Aromatic–Silica Hybrid Materials. *J. Mater. Chem.* **2005**, *15* (38), 4136–4140. https://doi.org/10.1039/B508818F.

(20) Morell, J.; Wolter, G.; Fröba, M. Synthesis and Characterization of Highly Ordered Thiophene-Bridged Periodic Mesoporous Organosilicas with Large Pores. *Chem. Mater.* **2005**, *17* (4), 804–808. https://doi.org/10.1021/cm048302d.

(21) Cho, E.-B.; Kim, D. Preparation of Highly Ordered Mesoporous Thiophene–Silica with Spherical Macrostructure. *Chem. Lett.* **2007**, *36* (1), 118–119. https://doi.org/10.1246/cl.2007.118.

(22) Álvaro, M.; Benítez, M.; Cabeza, J. F.; García, H.; Leyva, A. Electrochemiluminescence of a Periodic Mesoporous Organosilica Containing 9,10-Diarylanthracene Units. *J. Phys. Chem. C* **2007**, *111* (20), 7532–7538. https://doi.org/10.1021/jp0687394.

(23) Goto, Y.; Mizoshita, N.; Ohtani, O.; Okada, T.; Shimada, T.; Tani, T.; Inagaki, S. Synthesis of Mesoporous Aromatic Silica Thin Films and Their Optical Properties. *Chem. Mater.* **2008**, *20* (13), 4495–4498. https://doi.org/10.1021/cm800492s.

(24) Goto, Y.; Nakajima, K.; Mizoshita, N.; Suda, M.; Tanaka, N.; Hasegawa, T.; Shimada, T.; Tani, T.; Inagaki, S. Synthesis and Optical Properties of 2,6-Anthracene-Bridged Periodic Mesostructured Organosilicas. *Microporous Mesoporous Mater.* **2009**, *117* (3), 535–540. https://doi.org/10.1016/j.micromeso.2008.07.035.

(25) Mizoshita, N.; Tani, T.; Inagaki, S. Syntheses, Properties and Applications of Periodic Mesoporous Organosilicas Prepared from Bridged Organosilane Precursors. *Chem. Soc. Rev.* **2011**, *40* (2), 789–800. https://doi.org/10.1039/C0CS00010H.

(26) Park, M.; Park, S. S.; Selvaraj, M.; Zhao, D.; Ha, C.-S. Hydrophobic Mesoporous Materials for Immobilization of Enzymes. *Microporous Mesoporous Mater.* **2009**, *124* (1), 76–83. https://doi.org/10.1016/j.micromeso.2009.04.032.

(27) Huang, J.; Zhang, F. Periodic Mesoporous Organosilica Grafted Palladium Organometallic Complex: Efficient Heterogeneous Catalyst for Water-Medium Organic Reactions. *Appl. Organomet. Chem.* **2010**, *24* (11), 767–773. https://doi.org/10.1002/aoc.1696.

(28) Melero, J. A.; Bautista, L. F.; Morales, G.; Iglesias, J.; Sánchez-Vázquez, R. Acid-Catalyzed Production of Biodiesel over Arenesulfonic SBA-15: Insights into the Role of Water in the Reaction Network. *Renew. Energy* **2015**, *75*, 425–432. https://doi.org/10.1016/j.renene.2014.10.027.

(29) Bispo, C.; Ferreira, P.; Trouvé, A.; Batonneau-Gener, I.; Liu, F.; Jérôme, F.; Bion, N. Role of Acidity and Hydrophobicity in the Remarkable Catalytic Activity in Water of Sulfonic Acid-Functionalized Phenyl-PMO Materials. *Catal. Today* **2013**, *218–219*, 85–92. https://doi.org/10.1016/j.cattod.2013.06.004.

(30) Esquivel, D.; Jiménez-Sanchidrián, C.; Romero-Salguero, F. J. Thermal Behaviour, Sulfonation and Catalytic Activity of Phenylene-Bridged Periodic Mesoporous Organosilicas. *J. Mater. Chem.* **2010**, *21* (3), 724–733. https://doi.org/10.1039/C0JM02980G.

(31) Hall, D. A.; Maus, D. C.; Gerfen, G. J.; Inati, S. J.; Becerra, L. R.; Dahlquist, F. W.; Griffin, R. G. Polarization-Enhanced NMR Spectroscopy of Biomolecules in Frozen Solution. *Science* **1997**, *276* (5314), 930–932. https://doi.org/10.1126/science.276.5314.930.

(32) Lilly Thankamony, A. S.; Wittmann, J. J.; Kaushik, M.; Corzilius, B. Dynamic Nuclear Polarization for Sensitivity Enhancement in Modern Solid-State NMR. *Prog. Nucl. Magn. Reson. Spectrosc.* **2017**, *102–103*, 120–195. https://doi.org/10.1016/j.pnmrs.2017.06.002.

(33) Rankin, A. G. M.; Trébosc, J.; Pourpoint, F.; Amoureux, J.-P.; Lafon, O. Recent Developments in MAS DNP-NMR of Materials. *Solid State Nucl. Magn. Reson.* **2019**, *101*, 116–143. https://doi.org/10.1016/j.ssnmr.2019.05.009.

(34) Lesage, A.; Lelli, M.; Gajan, D.; Caporini, M. A.; Vitzthum, V.; Miéville, P.; Alauzun, J.; Roussey, A.; Thieuleux, C.; Mehdi, A.; et al. Surface Enhanced NMR Spectroscopy by Dynamic Nuclear Polarization. *J Am Chem Soc* **2010**, *132* (44), 15459–15461. https://doi.org/10.1021/ja104771z.

(35) Lafon, O.; Rosay, M.; Aussenac, F.; Lu, X.; Trébosc, J.; Cristini, O.; Kinowski, C.; Touati, N.; Vezin, H.; Amoureux, J.-P. Beyond the Silica Surface by Direct Silicon-29 Dynamic Nuclear Polarization. *Angew Chem Int Ed* **2011**, *50* (36), 8367–8370. https://doi.org/10.1002/anie.201101841.

(36) Liao, W.-C.; Ghaffari, B.; Gordon, C. P.; Xu, J.; Copéret, C. Dynamic Nuclear Polarization Surface Enhanced NMR Spectroscopy (DNP SENS): Principles, Protocols, and Practice. *Curr. Opin. Colloid Interface Sci.* **2018**, *33*, 63–71. https://doi.org/10.1016/j.cocis.2018.02.006.

(37) Grüning, W. R.; Rossini, A. J.; Zagdoun, A.; Gajan, D.; Lesage, A.; Emsley, L.; Copéret, C. Molecular-Level Characterization of the Structure and the Surface Chemistry of Periodic Mesoporous Organosilicates Using DNP-Surface Enhanced NMR Spectroscopy. *Phys. Chem. Chem. Phys. PCCP* **2013**, *15* (32), 13270–13274. https://doi.org/10.1039/c3cp00026e.

(38) Sánchez-Vázquez, R.; Pirez, C.; Iglesias, J.; Wilson, K.; Lee, A. F.; Melero, J. A. Zr-Containing Hybrid Organic–Inorganic Mesoporous Materials: Hydrophobic Acid Catalysts for Biodiesel Production. *ChemCatChem* **2013**, *5* (4), 994–1001. https://doi.org/10.1002/cctc.201200527.

(39) Sauvée, C.; Rosay, M.; Casano, G.; Aussenac, F.; Weber, R. T.; Ouari, O.; Tordo, P. Highly Efficient, Water-Soluble Polarizing Agents for Dynamic Nuclear Polarization at High Frequency. *Angew. Chem. Int. Ed.* **2013**, *125* (41), 11058–11061. https://doi.org/10.1002/ange.201304657.

(40) Rosay, M.; Tometich, L.; Pawsey, S.; Bader, R.; Schauwecker, R.; Blank, M.; Borchard, P. M.; Cauffman, S. R.; Felch, K. L.; Weber, R. T.; et al. Solid-State Dynamic Nuclear Polarization at 263 GHz: Spectrometer Design and Experimental Results. *Phys. Chem. Chem. Phys.* **2010**, *12* (22), 5850–5860. https://doi.org/10.1039/C003685B.

(41) Rossini, A. J.; Zagdoun, A.; Hegner, F.; Schwarzwälder, M.; Gajan, D.; Copéret, C.; Lesage, A.; Emsley, L. Dynamic Nuclear Polarization NMR Spectroscopy of Microcrystalline Solids. *J Am Chem Soc* **2012**, *134* (40), 16899–16908. https://doi.org/10.1021/ja308135r.

(42) Pinon, A. C.; Schlagnitweit, J.; Berruyer, P.; Rossini, A. J.; Lelli, M.; Socie, E.; Tang, M.; Pham, T.; Lesage, A.; Schantz, S.; et al. Measuring Nano- to Microstructures from Relayed

Dynamic Nuclear Polarization NMR. *J. Phys. Chem. C* **2017**, *121* (29), 15993–16005. https://doi.org/10.1021/acs.jpcc.7b04438.

(43) Fung, B. M.; Khitrin, A. K.; Ermolaev, K. An Improved Broadband Decoupling Sequence for Liquid Crystals and Solids. *J Magn Reson* **2000**, *142* (1), 97–101. https://doi.org/DOI: 10.1006/jmre.1999.1896.

(44) Pons, M.; Feliz, M.; Giralt, E. Steady-State DQF-COSY Spectra Using a Variable Relaxation Delay. *J Magn Reson* **1988**, *78*, 314–320.

(45) Maciel, G. E. Silica Surfaces: Characterization. In *Encyclopedia of Magnetic Resonance*; Harris, R. K., Wasylishen, R. E., Eds.; John Wiley & Sons, Ltd, 2007.

(46) Han, O.-H.; Baea, Y.-K. Solid-State NMR Study on the Structure and Dynamics of Triblock Copolymer P123 Remaining in SBA-15 after Solvent Washing. *Bull. Korean Chem. Soc.* **2008**, *29* (5), 911–912. https://doi.org/10.5012/bkcs.2008.29.5.911.

(47) Esquivel, D.; Jiménez-Sanchidrián, C.; Romero-Salguero, F. J. Thermal Behaviour, Sulfonation and Catalytic Activity of Phenylene-Bridged Periodic Mesoporous Organosilicas. *J. Mater. Chem.* **2011**, *21* (3), 724–733. https://doi.org/10.1039/C0JM02980G.

(48) Kuroki, M.; Asefa, T.; Whitnal, W.; Kruk, M.; Yoshina-Ishii, C.; Jaroniec, M.; Ozin, G. A. Synthesis and Properties of 1,3,5-Benzene Periodic Mesoporous Organosilica (PMO): Novel Aromatic PMO with Three Point Attachments and Unique Thermal Transformations. *J. Am. Chem. Soc.* **2002**, *124* (46), 13886–13895. https://doi.org/10.1021/ja027877d.

(49) Onida, B.; Borello, L.; Busco, C.; Ugliengo, P.; Goto, Y.; Inagaki, S.; Garrone, E. The Surface of Ordered Mesoporous Benzene−Silica Hybrid Material: An Infrared and Ab Initio

Molecular Modeling Study. *J. Phys. Chem. B* **2005**, *109* (24), 11961–11966. https://doi.org/10.1021/jp050686n.

(50) Camarota, B.; Onida, B.; Goto, Y.; Inagaki, S.; Garrone, E. Hydroxyl Species in Large-Pore Phenylene-Bridged Periodic Mesoporous Organosilica. *Langmuir* **2007**, *23* (26), 13164–13168. https://doi.org/10.1021/la702252j.

(51) Styskalik, A.; Skoda, D.; Moravec, Z.; Roupcova, P.; Barnes, C. E.; Pinkas, J. Non-Aqueous Template-Assisted Synthesis of Mesoporous Nanocrystalline Silicon Orthophosphate. *RSC Adv.* **2015**, *5* (90), 73670–73676. https://doi.org/10.1039/C5RA10982E.

(52) Chandra, D.; Yokoi, T.; Tatsumi, T.; Bhaumik, A. Highly Luminescent Organic−Inorganic Hybrid Mesoporous Silicas Containing Tunable Chemosensor inside the Pore Wall. *Chem. Mater.* **2007**, *19* (22), 5347–5354. https://doi.org/10.1021/cm701918t.

(53) Jones, L. H.; McLaren, E. Infrared Spectra of CH3COONa and CD3COONa and Assignments of Vibrational Frequencies. *J. Chem. Phys.* **1954**, *22* (11), 1796–1800. https://doi.org/10.1063/1.1739921.

(54) Oomens, J.; Steill, J. D. Free Carboxylate Stretching Modes. *J. Phys. Chem. A* **2008**, *112* (15), 3281–3283. https://doi.org/10.1021/jp801806e.

(55) Lei, L.; Shan, J.; Hu, J.; Liu, X.; Zhao, J.; Tong, Z. Co-Curing Effect of Imidazole Grafting Graphene Oxide Synthesized by One-Pot Method to Reinforce Epoxy Nanocomposites. *Compos. Sci. Technol.* **2016**, *128*, 161–168. https://doi.org/10.1016/j.compscitech.2016.03.029.

(56) Li, K.-Y.; Kuan, C.-F.; Kuan, H.-C.; Chen, C.-H.; Shen, M.-Y.; Yang, J.-M.; Chiang, C.-L. Preparation and Properties of Novel Epoxy/Graphene Oxide Nanosheets (GON) Composites

Functionalized with Flame Retardant Containing Phosphorus and Silicon. *Mater. Chem. Phys.* **2014**, *146* (3), 354–362. https://doi.org/10.1016/j.matchemphys.2014.03.037.

(57) Hwang, Y.; Heo, Y.; Yoo, Y.; Kim, J. The Addition of Functionalized Graphene Oxide to Polyetherimide to Improve Its Thermal Conductivity and Mechanical Properties. *Polym. Adv. Technol.* **2014**, *25* (10), 1155–1162. https://doi.org/10.1002/pat.3369.

(58) Ramasamy, R. Vibrational Spectroscopic Studies of Imidazole. *Armen. J. Phys.* **2015**, *8*, 51–55.

(59) Kryszak, D.; Stawicka, K.; Calvino-Casilda, V.; Martin-Aranda, R.; Ziolek, M. Imidazole Immobilization in Nanopores of Silicas and Niobiosilicates SBA-15 and MCF—A New Concept towards Creation of Basicity. *Appl. Catal. Gen.* **2017**, *531*, 139–150. https://doi.org/10.1016/j.apcata.2016.10.028.

(60) Graf, N.; Yegen, E.; Gross, T.; Lippitz, A.; Weigel, W.; Krakert, S.; Terfort, A.; Unger, W. E. S. XPS and NEXAFS Studies of Aliphatic and Aromatic Amine Species on Functionalized Surfaces. *Surf. Sci.* **2009**, *603* (18), 2849–2860. https://doi.org/10.1016/j.susc.2009.07.029.

(61) Truica-Marasescu, F.; Wertheimer, M. R. Nitrogen-Rich Plasma-Polymer Films for Biomedical Applications. *Plasma Process. Polym.* **2008**, *5* (1), 44–57. https://doi.org/10.1002/ppap.200700077.

TOC

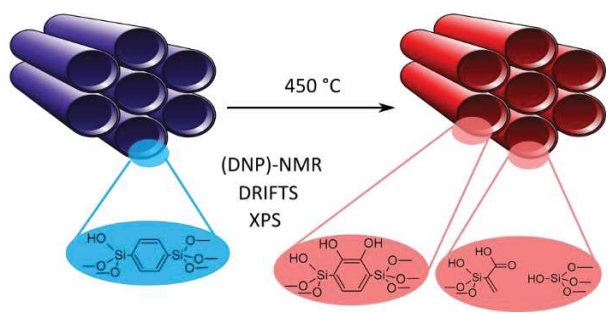

# Probing functionalities and acidity of calcined phenylene-bridged periodic mesoporous organosilicates using (DNP)-NMR, DRIFTS and XPS

## Supplementary information

*Cyril Pirez[a*], Hiroki Nagashima[b], Franck Dumeignil[a], Olivier Lafon[a,c*]*

[a] Univ. Lille, CNRS, Centrale Lille, ENSCL, Univ. Artois, UMR 8181 – UCCS – Unité de Catalyse et Chimie du Solide, F-59000 Lille, France
[b] Interdisciplinary Research Center for Catalytic Chemistry, National Institute of Advanced Industrial Science and Technology (AIST), 1-1-1 Higashi, Tsukuba, Ibaraki 305-8565, Japan
[c] Institut Universitaire de France (IUF), 1 rue Descartes, 75231 Paris, France

[*] Corresponding author. Tel.: +33 (0)3 20 33 77 35; fax: +33 (0)3 20 33 65 61. E-mail address: cyril.pirez@gmail.com, olivier.lafon@univ-lille.fr

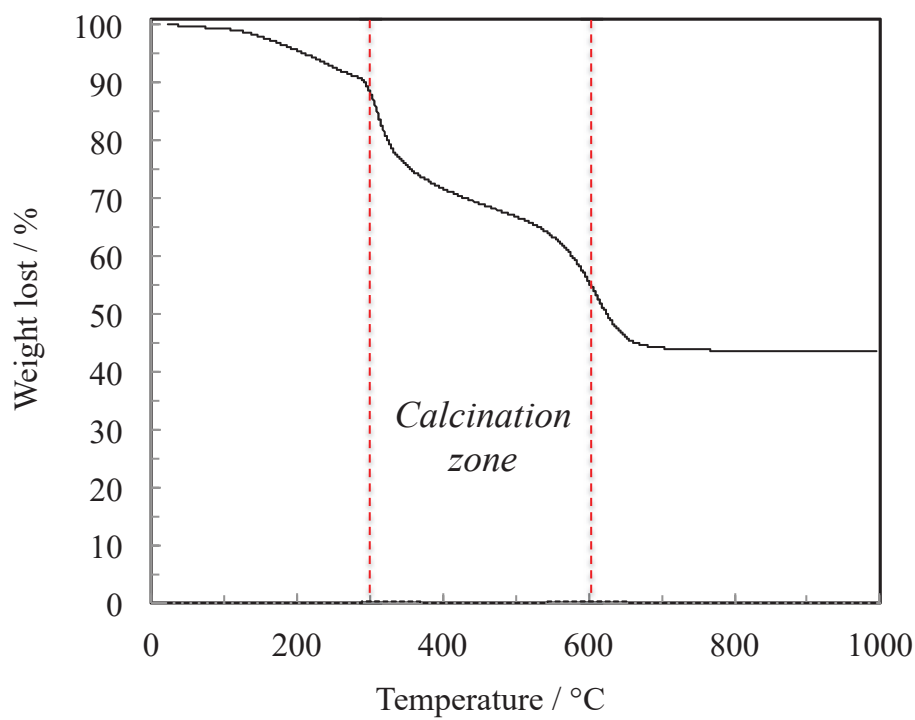

**Figure S1**. Thermogravimetry analysis of as-synthesized PMOB material containing template.

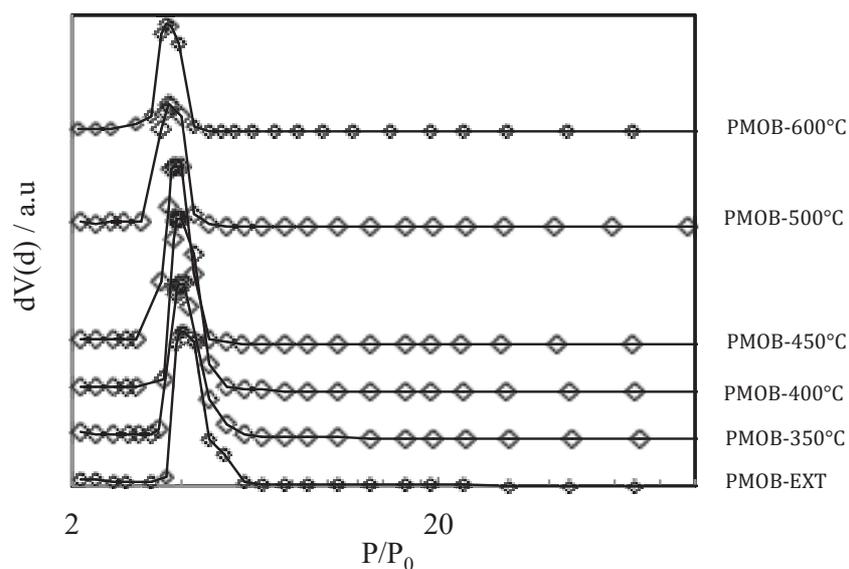

**Figure S2.** Pore size distribution of PMOB materials.

**Table S1.** Structural and textural properties of PMOB materials. $V_p$ and $W_{BJH}$ denotes the porous volume and the pore diameter, respectively.

| Catalyst | Calcination / °C | BET / m².g$^{-1}$ | $V_p$ / cm³·g$^{-1}$ | $W_{BJH,DES}$ / nm | Plane spacing / nm | Unit cell parameter / nm | Wall thickness / nm |
|---|---|---|---|---|---|---|---|
| PMOB-EXT | 25 | 713 | 0.66 | 4.1 | 10.5 | 12.1 | 8.0 |
| PMOB-350C | 350 | 986 | 0.87 | 4 | 9.8 | 11.3 | 7.3 |
| PMOB-400C | 400 | 709 | 0.63 | 3.9 | 9.4 | 10.8 | 7.0 |
| PMOB-450C | 450 | 556 | 0.5 | 3.8 | 8.7 | 10.0 | 6.4 |
| PMOB-500C | 500 | 472 | 0.43 | 3.7 | 7.5 | 8.6 | 5.0 |
| PMOB-600C | 600 | 374 | 0.34 | 3.6 | 7.2 | 8,4 | 4.7 |

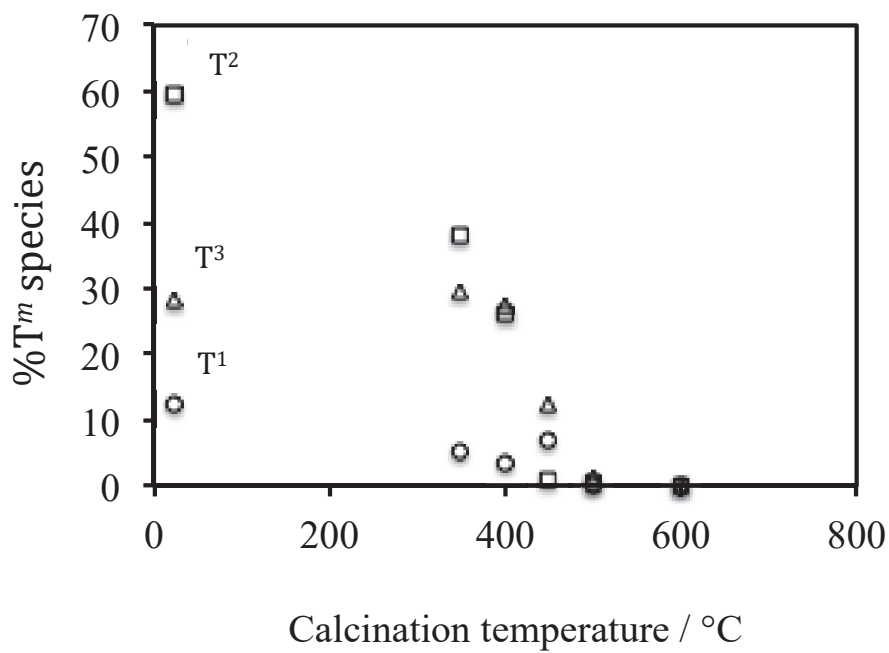

**Figure S3.** Plot of the $^{29}$Si signal fractions corresponding to T$^m$ sites in PMOB materials as function of the calcination temperature. The signal fractions correspond to the relative integrated intensities of the different T$^m$ sites in the 1D $^{29}$Si NMR spectrum of Fig. 4.

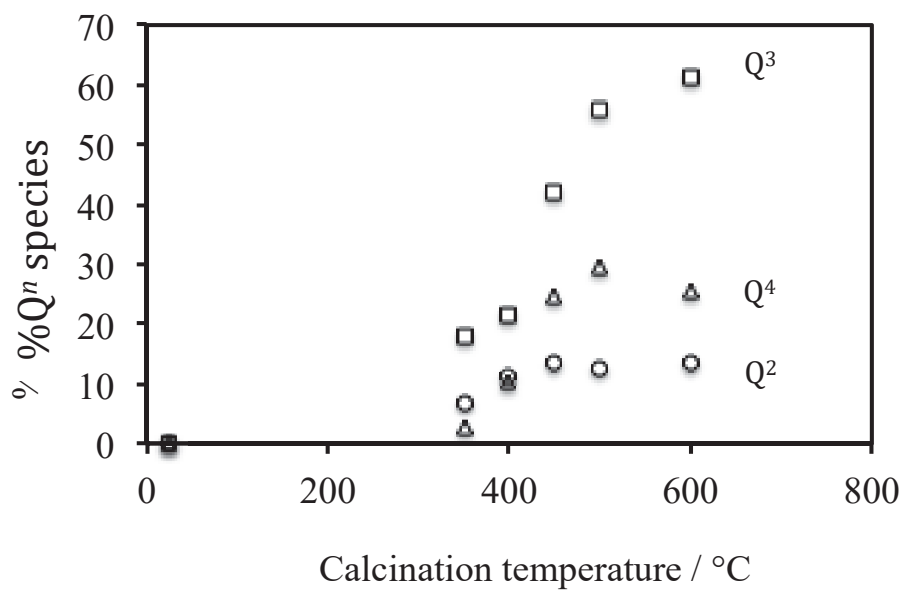

**Figure S4.** Plot of the $^{29}$Si signal fractions corresponding to Q$^n$ sites in PMOB materials as function of the calcination temperature. The signal fractions correspond to the relative integrated intensities of the different Q$^n$ sites in the 1D $^{29}$Si NMR spectrum of Fig. 4.

**Table S2**. $^{29}$Si signal fractions corresponding to $T^m$ and $Q^n$ sites as function of calcination temperature. The signal fractions correspond to the relative integrated intensities of the different $Q^n$ sites in the 1D $^{29}$Si NMR spectrum of Fig. 4.

| Calcination temperature /°C | $T^m$ species / % | | | $Q^n$ species / % | | |
|---|---|---|---|---|---|---|
| | $T^1$ | $T^2$ | $T^3$ | $Q^2$ | $Q^3$ | $Q^4$ |
| 25 | 12.3 | 59.4 | 28.3 | 0 | 0 | 0 |
| 350 | 5.2 | 38.2 | 29.5 | 6.5 | 18.0 | 2.6 |
| 400 | 3.4 | 26.1 | 27.4 | 11.2 | 21.5 | 10.5 |
| 450 | 6.7 | 0.9 | 12.3 | 13.3 | 42.2 | 24.6 |
| 500 | 0 | 0.6 | 1.5 | 12.6 | 56.0 | 29.4 |
| 600 | 0 | 0 | 0 | 13.3 | 61.3 | 25.4 |

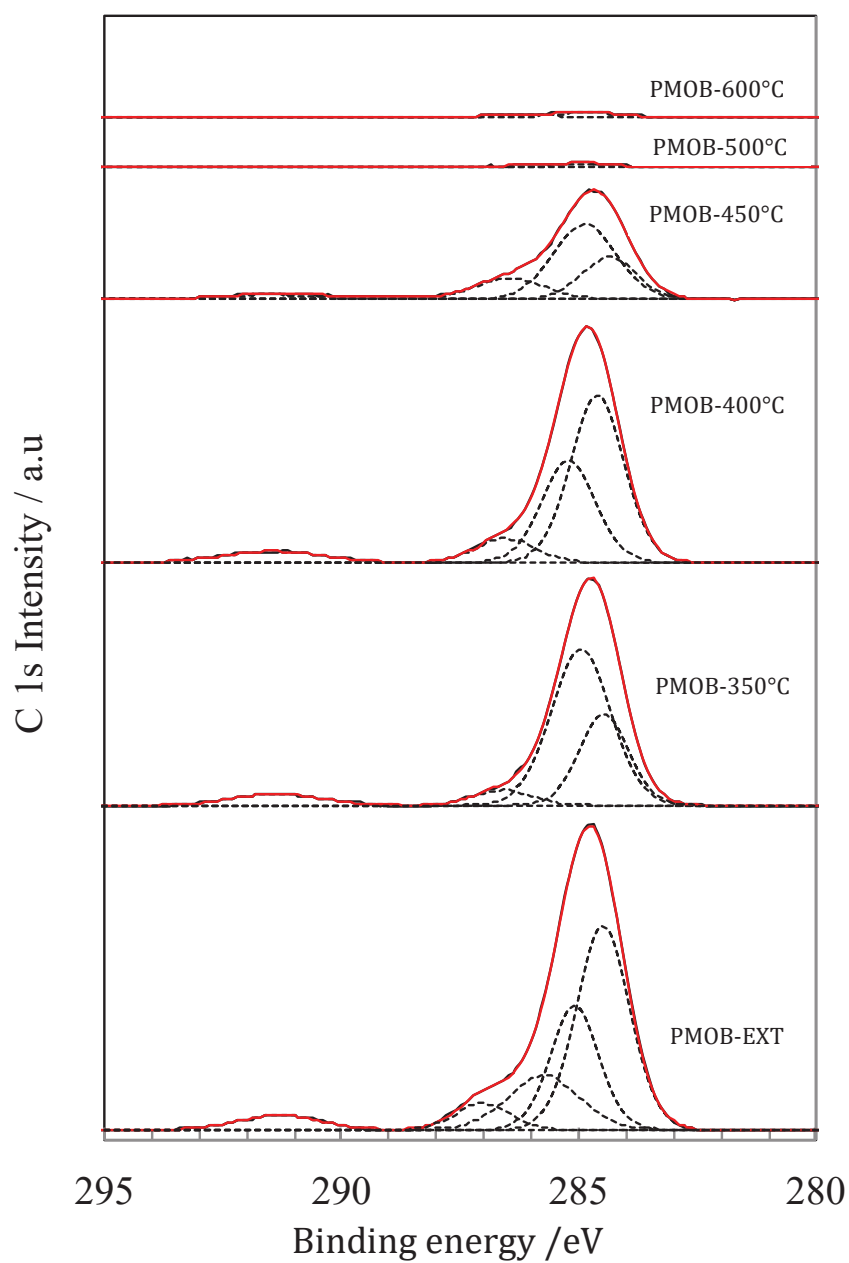

**Figure S5**. 1s C XPS spectra of PMOB-EXT and PMOB calcined.

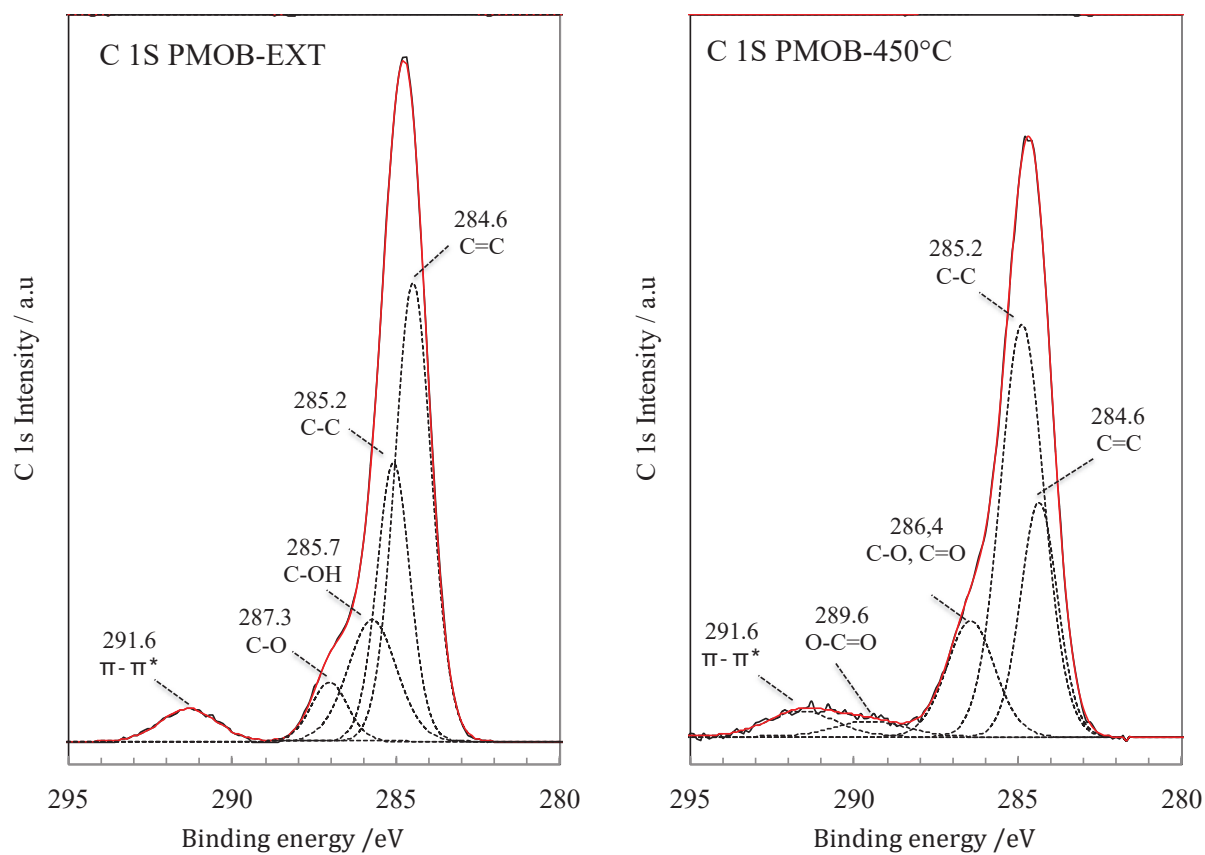

**Figure S6**. 1s C XPS spectra of PMOB-EXT and PMOB-450°C.

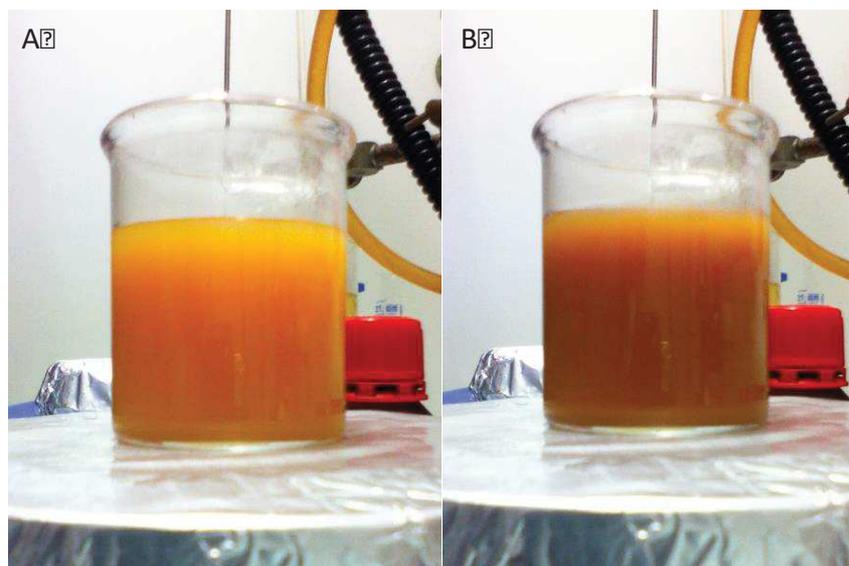

**Figure S7**. Color change of (A) a dispersion of PMOB-450°C particles in water after (B) the addition of one drop of 1-(3-aminopropyl)imidazole (colorless product).

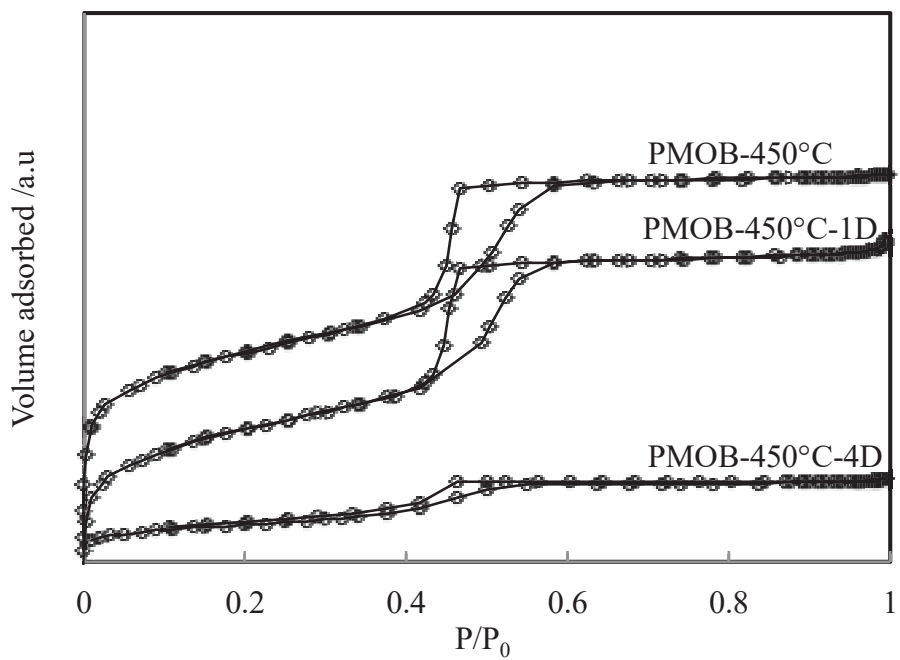

**Figure S8**. N$_2$ adsorption-desorption of the PMOB-450°C material after addition of pure 1-(3-aminopropyl)imidazole in distilled water at room temperature.

**Table S3**. Surface area, pore volume and pore diameter of PMOB-450°C materials after addition of pure API.

| Catalyst | Weight / mg | Number of API drops | BET / m².g | V$_p$ / cm$^3$·g | W$_{BJH\,ads}$ / nm | W$_{BJH\,des}$ / nm |
|---|---|---|---|---|---|---|
| PMOB-450°C | - | - | 524 | 0.43 | 3.6 | 3.6 |
| PMOB-450°C-1D | 500 | 1 | 377 | 0.37 | 3.7 | 3.5 |
| PMOB-450°C-4D | 500 | 4 | 42 | 0.04 | 3.2 | 3.4 |

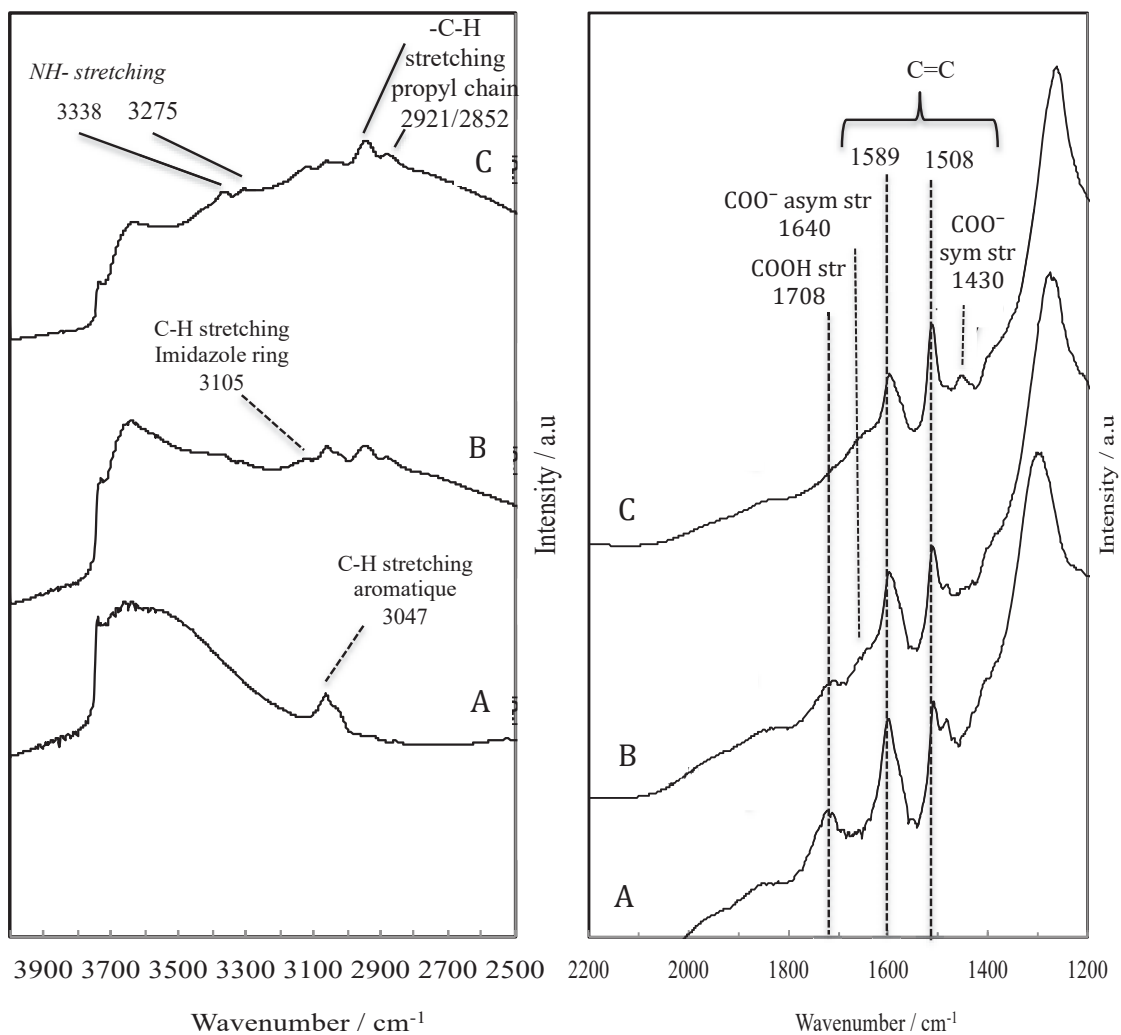

**Figure S9.** Expansion of the (left) 4000-2500 cm$^{-1}$ and (right) 2000-1000 cm$^{-1}$ regions of DRIFTS spectra of (A) PMOB-450°, (B) PMOB-450°C-1D and (C) PMOB-450°C-4D.

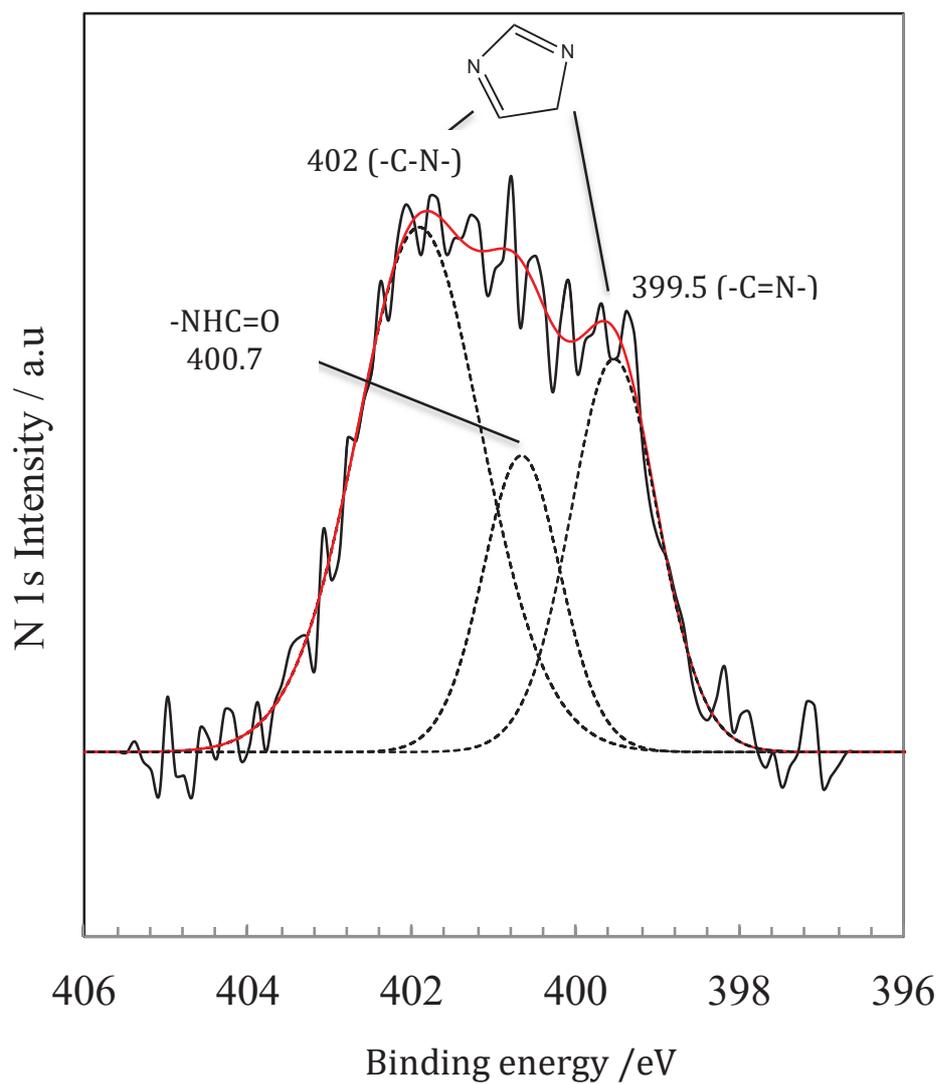

**Figure S10**. 1s N XPS spectra of PMO-450°C after API grafting by covalent binding.